%% file: main.tex
\documentclass[twocolumn]{aastex631}
\usepackage{graphicx}
\usepackage{xcolor}
\usepackage{amsmath} 
\usepackage{amssymb}
\usepackage{rotating}
\usepackage{url}
\usepackage{comment}
\usepackage{enumitem}

\shorttitle{Multi-wavelength view of the close-by GRB~190829A}
\shortauthors{Salafia et al.}

\begin{document}

\title{Multi-wavelength view of the close-by GRB~190829A sheds light on gamma-ray burst physics}

\author[0000-0003-4924-7322]{Om Sharan Salafia}
\affiliation{INAF -- Osservatorio Astronomico di Brera, via E. Bianchi 46, I-23807 Merate (LC), Italy}
\affiliation{INFN -- Sezione di Milano-Bicocca, piazza della Scienza 2, I-20126 Milano (MI), Italy}
\affiliation{Università degli Studi di Milano-Bicocca, Dip.\ di Fisica ``G. Occhialini'', piazza della Scienza 3, I-20126 Milano (MI), Italy}
 
\author[0000-0003-3193-4714]{Maria Edvige Ravasio}
\affiliation{Università degli Studi di Milano-Bicocca, Dip.\ di Fisica ``G. Occhialini'', piazza della Scienza 3, I-20126 Milano (MI), Italy}
\affiliation{INAF -- Osservatorio Astronomico di Brera, via E. Bianchi 46, I-23807 Merate (LC), Italy}

\author{Jun Yang}
\affiliation{Department of Space, Earth and Environment, Chalmers University of Technology, Onsala Space Observatory, SE-439 92 Onsala, Sweden}

\author{Tao An}
\affiliation{Shanghai Astronomical Observatory, Key Laboratory of Radio Astronomy, Chinese Academy of Sciences, Nandan Road 80, 200030, China}
\affiliation{Key Laboratory of Cognitive Radio and Information Processing, Guilin University of Electronic Technology, 541004, Guilin, China}

\author{Monica Orienti} 
\affiliation{INAF -- Istituto di Radioastronomia, via P. Gobetti 101, I-40129 Bologna, Italy}

\author{Giancarlo Ghirlanda} 
\affiliation{INAF -- Osservatorio Astronomico di Brera, via E. Bianchi 46, I-23807 Merate (LC), Italy}
\affiliation{INFN -- Sezione di Milano-Bicocca, piazza della Scienza 2, I-20126 Milano (MI), Italy}

\author{Lara Nava}
\affiliation{INAF -- Osservatorio Astronomico di Brera, via E. Bianchi 46, I-23807 Merate (LC), Italy}
\affiliation{INFN -- Sezione di Trieste, via Valerio 2, I-34127 Trieste (TS), Italy}

\author{Marcello Giroletti} 
\affiliation{INAF -- Istituto di Radioastronomia, via P. Gobetti 101, I-40129 Bologna, Italy}

\author{Prashanth Mohan}
\affiliation{Key Laboratory of Cognitive Radio and Information Processing, Guilin University of Electronic Technology, 541004, Guilin, China}

\author{Riccardo Spinelli}  
\affiliation{Dipartimento di Scienza e Alta Tecnologia, Universit\`a dell'Insubria, Via Valleggio 11, 22100 Como, Italy}
\affiliation{INAF -- Osservatorio Astronomico di Brera, via E. Bianchi 46, I-23807 Merate (LC), Italy}

\author{Yingkang Zhang}
\affiliation{Shanghai Astronomical Observatory, Key Laboratory of Radio Astronomy, Chinese Academy of Sciences, Nandan Road 80, 200030, China}

\author{Benito Marcote} 
\affiliation{Joint Institute for VLBI ERIC, Oude Hoogeveensedijk 4, 7991~PD Dwingeloo, The Netherlands}

\author{Giuseppe Cim\`{o}}
\affiliation{Joint Institute for VLBI ERIC, Oude Hoogeveensedijk 4, 7991~PD Dwingeloo, The Netherlands}

\author{Xuefeng Wu} 
\affiliation{Purple Mountain Observatory, Chinese Academy of Sciences, Nanjing 210023, China}

\author{Zhixuan Li}
\affiliation{Yunnan Observatories, Chinese Academy of Sciences, 650216 Kunming, Yunnan, China}

\begin{abstract} 
We monitored the position of the close-by (about 370 Mpc) gamma-ray burst GRB 190829A, which originated from a massive star collapse, through very long baseline interferometry (VLBI) observations with the European VLBI Network and the Very Long Baseline Array, carrying out a total of 9 observations between 9 and 117 days after the gamma-ray burst at 5 and 15 GHz, with a typical resolution of few mas. From a state-of-the art analysis of these data, we obtained valuable limits on the source size and expansion rate. The limits are in agreement with the size evolution entailed by a detailed modelling of the multi-wavelength light curves with a forward plus reverse shock model, which agrees with the observations across almost 18 orders of magnitude in frequency (including the HESS data at TeV photon energies) and more than 4 orders of magnitude in time. Thanks to the multi-wavelength, high-cadence coverage of the afterglow, inherent degeneracies in the afterglow model are broken to a large extent, allowing us to capture some unique physical insights: we find a low prompt emission efficiency $\lesssim 10^{-3}$; a low fraction of relativistic electrons in the forward shock downstream $\chi_e<13\%$ (90\% credible level); a rapid decay of the magnetic field in the reverse shock downstream after the shock crossing. While our model assumes an on-axis jet, our VLBI astrometry is not sufficiently tight as to exclude any off-axis viewing angle, but we can exclude the line of sight to have been more than $\sim 2\,\mathrm{deg}$ away from the border of the gamma-ray-producing region based on compactness arguments.
\end{abstract}

\section{Introduction}

Radio observations of gamma-ray burst (GRB) afterglows have been rarely successful in constraining their projected size or proper motion due to the large distances involved. In a handful of cases \citep{Taylor1998,Taylor1999,Frail2000,Alxander2019}, such as that of GRB~970508 \citep{Frail1997}, scintillation of the radio source induced by scattering of the emission by the interstellar medium has been used as an indirect probe of the source size. On the other hand, the only case so far in which Very Long Baseline Interferometry (VLBI) observations could produce a direct measurement of the size of a GRB afterglow is that of GRB~030329 \citep{Taylor2004}. More recently, VLBI observations of GRB~170817A \citep{Mooley2018,Ghirlanda2019} led to direct inference of the effects of relativistic motion, that is, an apparently superluminal displacement of the source centroid. In these favourable cases, the joint modelling of the light curves and of the evolution of the apparent size \citep{Mesler2013} or the centroid displacement \citep{Ghirlanda2019,Hotokezaka2019} helped to mitigate the problem of afterglow model degeneracies, which most often prevents the determination of the source's physical properties unless some parameters are fixed based on educated guesses.

At the other end of the electromagnetic spectrum, observations of GRB afterglows at teraelectronvolt (TeV) photon energies \citep{Zhang2001,Nava2018} have also shown a potential in breaking the modelling degeneracies and constrain the underlying physical processes. Such photon energies are in principle beyond the reach of synchrotron emission from shock-accelerated electrons \citep{DeJager1992,Nava2018,HESS2021}: inverse Compton scattering of the synchrotron photons by the same relativistic electrons (`synchrotron self-Compton', \cite{Rybicki1986,Panaitescu1998,Chiang1999,Panaitescu2000,Sari2001}) is expected to dominate at these energies. Such process was shown to provide a viable explanation \citep{MAGIC2019IC} for the TeV emission component recently detected \citep{MAGIC2019OBS} by the Major Atmospheric Gamma Imaging Cherenkov (MAGIC) in association to GRB~190114C. Different emission processes mean different dependencies on the physical properties of the source, which enhances the prospects for breaking the degeneracies.  Unfortunately, TeV observations of gamma-ray bursts are notoriously difficult, and only few detections have been reported so far \citep{Atkins2000,MAGIC2019OBS,HESS201920B,GCN_201015A,GCN_201216C}, including the source \citep{HESS2019GCN,HESS2021} we study in this work.

\section{Results}

\subsection{The GRB~190829A event}

GRB~190829A is a long GRB detected by the Gamma-ray Burst Monitor (GBM) onboard the \textit{Fermi} satellite on 2019 August 29 at 19:55:53 UT \citep{2019GCN.25551Fermi} and shortly thereafter \citep{2019GCN.25552Swift} by the Burst Alert Telescope (BAT) onboard the Neil Gehrels Swift Observatory (\textit{Swift} hereafter). 
GRB~190829A is the third GRB detected \cite{HESS2019GCN} at teraelectronvolt photon energies after GRB~190114C \citep{MAGIC2019} and GRB~180720B \citep{HESS2019}, but, compared to these, it features a smaller isotropic-equivalent energy \citep[][$E_\mathrm{iso} \sim 3 \times 10^{50}$ erg -- see also Appendix~\ref{sec:GBM_reduction}]{2019GCN.25660KonusWind}.
The redshift of the host galaxy $z = 0.0785$ (\citealt{Valeev2019}, corresponding to a luminosity distance of approximately $368$ Mpc adopting Planck cosmological parameters -- \citealt{Planck2016} -- or equivalently an angular diameter distance of $316$ Mpc) makes this event one of the closest long GRBs known to date. 
The afterglow emission of GRB~190829A has been monitored up to several months after the burst: after an initial peak and a fading phase, a re-brightening in the optical light curve at $\sim$5 days was attributed to the associated supernova emission (confirmed by the spectroscopic observations of the 10.4m Gran Telescopio Canarias telescope, hereafter GTC -- \citealt{Hu2020}).
Radio afterglow emission was first detected by the Australia Telescope Compact Array (ATCA) at 5.5 GHz \citep{Laskar2019} and then by the Northern Extended Millimeter Array (NOEMA) at 90 GHz \citep{2019GCN.25589NOEMA}, 20.2 hours and 29.48 hours after the burst, respectively. Subsequent high-cadence radio observations were performed with the Meer Karoo Array Telescope (MeerKAT) at 1.3 GHz and Arcminute Microkelvin Imager--Large Array (AMI-LA) at 15.5 GHz, reporting a fading radio source up to 143 days after the initial gamma-ray burst \citep{Rhodes2020}.

\subsection{VLBI observations and Sedov length constraint} 

\begin{figure}[t]
\includegraphics[width=\columnwidth]{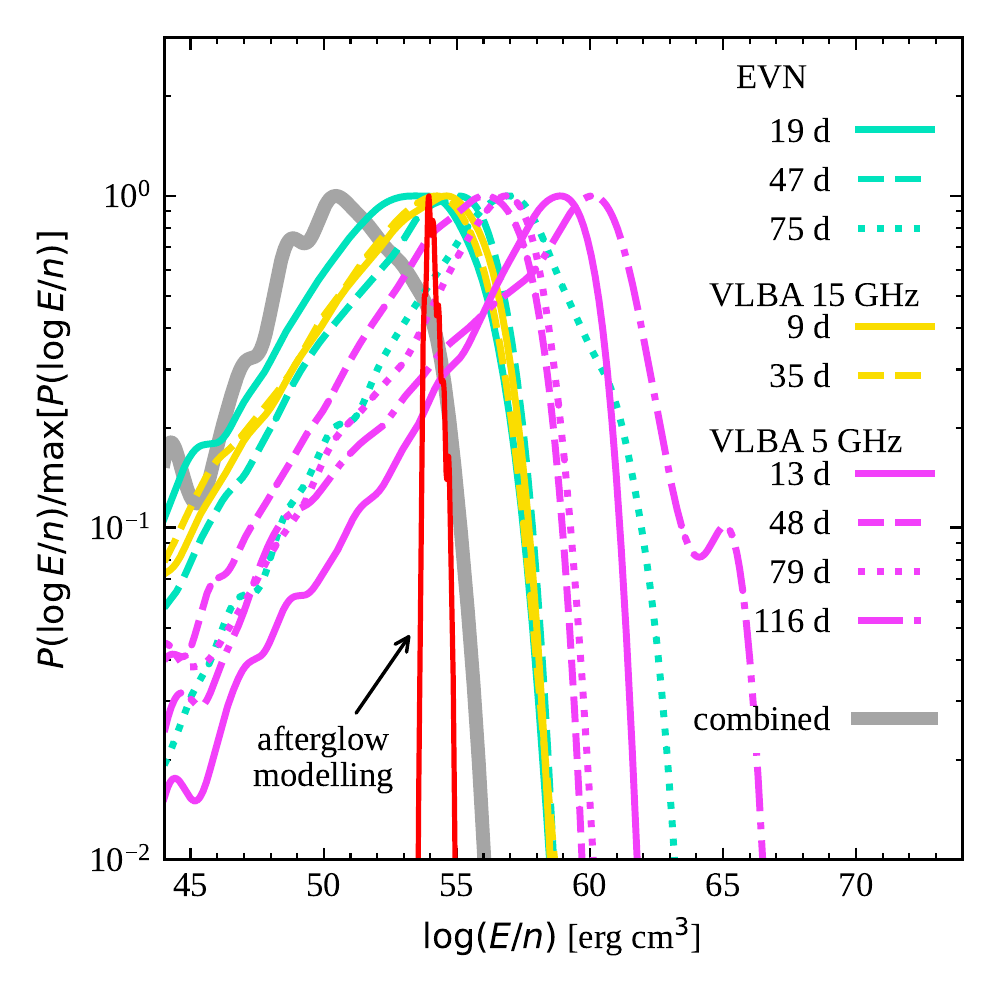}
\caption{Constraint on the energy-to-density ratio. Turquoise, orange and fuchsia lines show the posterior probability on $\log(E/n)$ obtained (Appendix~\ref{sec:sedov_length_constraint_method}) from the source size measurements (Appendix~\ref{sec:vlbi_source_fit_method}) in our VLBI imaging epochs assuming the source to be a relativistic shock in self-similar expansion \citep{Blandford1976,Granot1999}. The grey line shows the combined posterior from all epochs. The red solid line is the posterior obtained from fitting our forward plus reverse shock afterglow emission model to the available multi-wavelength data.}
\label{fig:E_n_constraint}
\end{figure}

We conducted VLBI observations of GRB~190829A with the Very Long Baseline Array (VLBA) at 15 and 5 GHz and the European VLBI Network (EVN) alongside the enhanced Multi-Element Remotely Linked Interferometer Network (e-MERLIN) at 5 GHz, for a total of nine epochs between 9 and 116 days after the GRB (see Table \ref{tab:obs}).
\begin{figure*}[t]
\includegraphics[width=\textwidth]{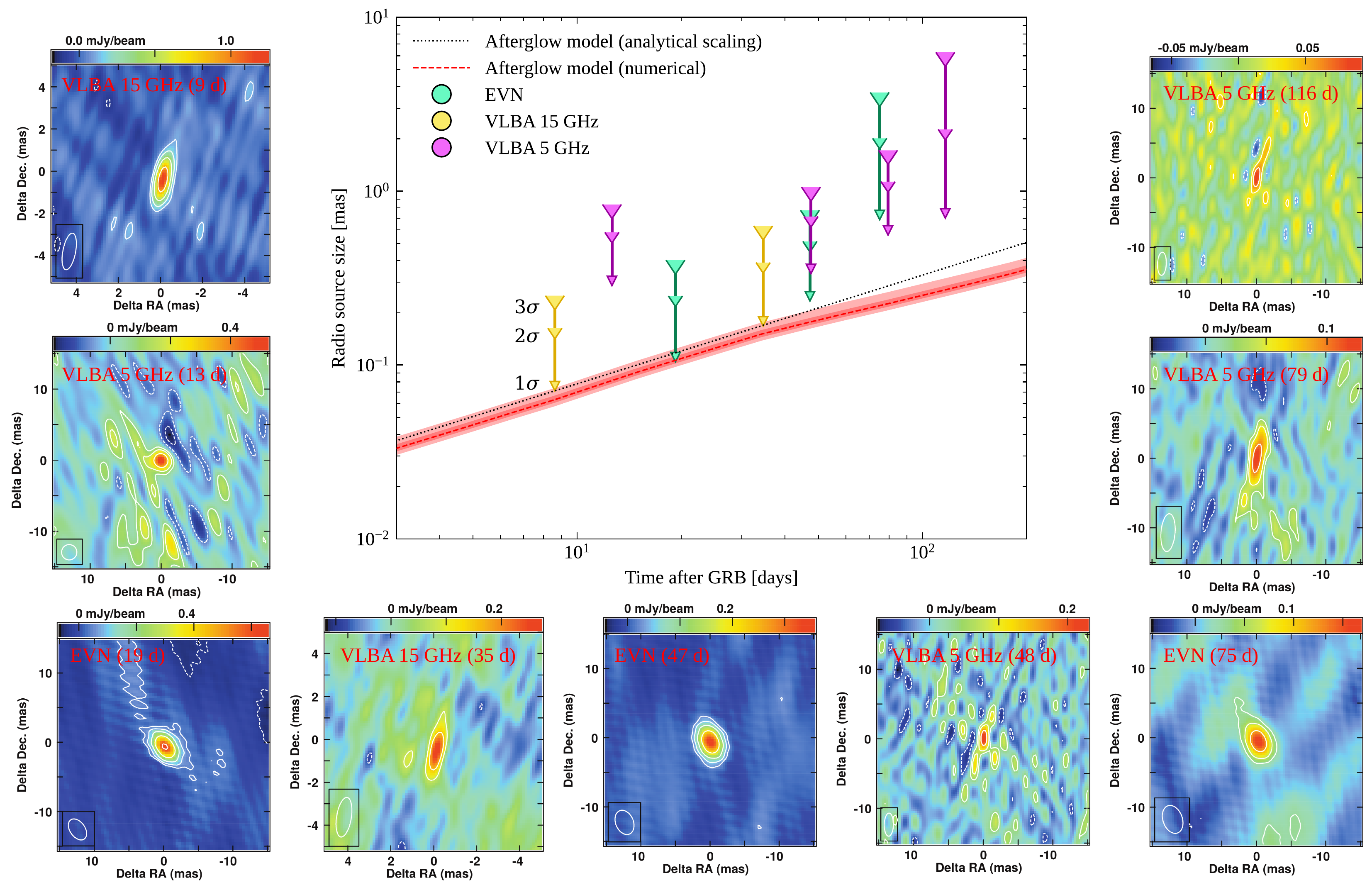}
\caption{Source size upper limits and comparison with the model. Central panel: Downward arrows show our one-, two- and three-sigma upper limits on the source size at each epoch (see Appendix~\ref{sec:sedov_length_constraint_method}). The dashed lines show the source size evolution as predicted by our afterglow model (analytical $t^{5/8}$ scaling -- \citealt{Granot1999} -- in black, source size from our numerical model in red) assuming our best-fit parameters (see Appendix~\ref{sec:afterglow_model_fitting}). The pink shaded band shows the 90\% credible interval implied by our afterglow parameter uncertainties. The surrounding smaller panels show previews of the cleaned radio maps for each epoch (full-size maps are \textbf{available on Zenodo} -- see \citealt{zenodo_supplementary}).  
}
\label{fig:size_evolution}
\end{figure*}
Despite the good angular resolution reached in all observations, the source remained consistently unresolved. In order to obtain reliable upper limits on the source size, we fitted a circular Gaussian model to the data through a Markov Chain Monte Carlo approach (Appendix~\ref{sec:vlbi_source_fit_method}), which we first tested against simulated sources immersed in real noise (Appendix~\ref{sec:vlbi_source_fit_validation}). From the analysis of our nine-epoch data, we obtained the limits reported in Table~\ref{tab:fit} and shown in Figure~\ref{fig:size_evolution}. Assuming an intrinsic source size evolution $s\propto t^{5/8}$ as expected \citep{Granot1999} for the observed size $s$ of a relativistic blastwave whose expansion is described by the self-similar Blandford-McKee solution \citep{Blandford1976}, we could translate our measurements into a largely model-independent upper limit on the ratio between the blastwave energy $E$ and the number density $n$ of the surrounding ambient medium, which sets the fundamental length scale of the expansion, namely the Sedov length \citep{Blandford1976} $\ell_\mathrm{S}=(3E/4\pi n m_\mathrm{p} c^2)^{1/3}$, where $m_\mathrm{p}$ is the proton mass and $c$ is the speed of light. Since \citep{Blandford1976,Granot1999} $s\propto \ell_\mathrm{S}^{3/8} t^{5/8}$, we have that $E/n \propto \ell_\mathrm{S}^3 \propto s^8 t^{-5}$. After carefully evaluating the proportionality constant (Appendix~\ref{sec:sedov_length_constraint_method}) and adopting a flat prior on the source size, we obtained the posterior probabilities shown in Figure~\ref{fig:E_n_constraint}. We note that, since we do not resolve the source, only upper limits derived from these posteriors are meaningful. The most stringent upper limit is that from our first EVN epoch (solid turquoise line in Fig.~\ref{fig:E_n_constraint}), which yielded $\log[(E/n)/\mathrm{erg\,cm^{3}}] < 55.6$ at the 90\% credible level. After combining the posterior probabilities from all the epochs (grey thick line in Fig.~\ref{fig:E_n_constraint}, see Appendix~\ref{sec:sedov_length_constraint_method}) we obtained $\log[(E/n)/\mathrm{erg\,cm^{3}}] < 54.1$ at the 90\% credible level.

\subsection{Time-dependent multi-wavelength modelling and interpretation}
In order to test this result and get a deeper physical insight on this source, we performed a self-consistent modelling of all the available multi-wavelength observations of the afterglow. We included both the forward and reverse shock emission in our model, assuming a uniform angular profile for all jet properties within an opening angle $\theta_\mathrm{j}$ and computing the shock dynamics self-consistently from deceleration down to the late side expansion phase. We computed the radiation in the shock downstream comoving frame including the effects of inverse Compton scattering on electron cooling (accounting for the Klein-Nishina suppression of the cross section above the relevant photon energy), assuming a fixed fraction $\epsilon_\mathrm{e}$ of the available energy density to be in relativistic electrons (which we assumed to be a fraction $\chi_\mathrm{e}$ of the total electrons, and to be injected with a power law energy distribution with index $p>2$), and a fraction $\epsilon_\mathrm{B}$ to be in the form of an effectively isotropic magnetic field. To compute the observed emission, we integrated over equal-arrival-time surfaces and considered relativistic beaming effects.\\
 
\begin{figure*}
\centering
\includegraphics[width=\textwidth]{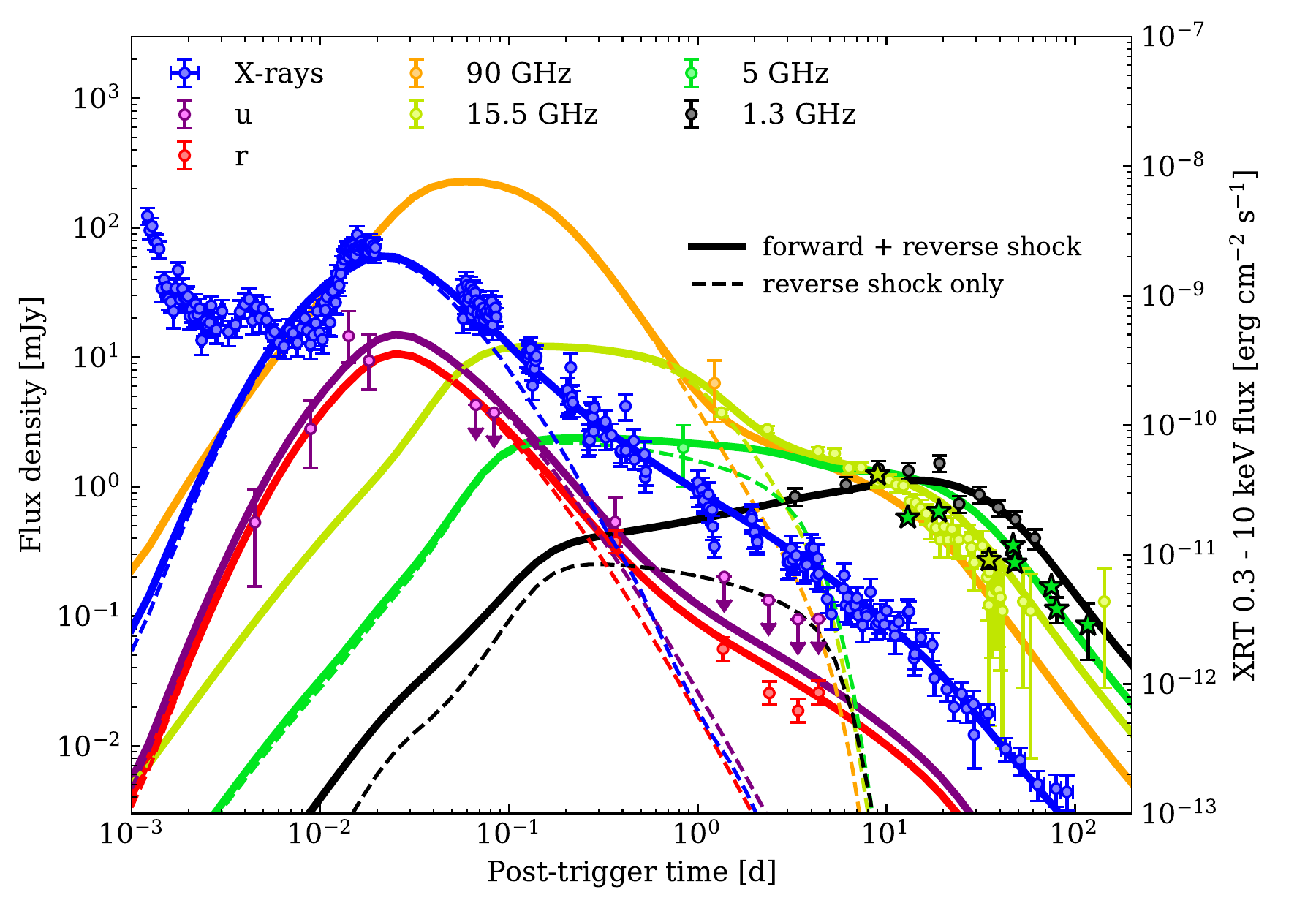}    

\caption{Multi-wavelength data and emission model. Circles represent X-ray fluxes (blue, values shown on the right-hand axis) or flux densities (all other colours, values shown on the left-hand axis) measured at the position of GRB190829A at different times after the GRB trigger in several bands (see the legend). Optical flux densities have been corrected for both the Milky Way and host galaxy extinction, and the contribution of the host galaxy has been subtracted. The host galaxy contribution \citep{Rhodes2020} has also been subtracted from the AMI-LA radio flux densities at 15.5 GHz. Stars mark the flux densities measured in our VLBI epochs. Solid lines of the corresponding colours show the predictions of our emission model including both the forward and reverse shocks. Dashed lines single out the contribution of the reverse shock emission. We interpret the initial plateau in the X-ray data as the superposition of the prompt emission tail and the rising reverse shock emission.
}
\label{fig:lightcurves}
\end{figure*}

\begin{figure*}[t]
\begin{center}
\includegraphics[width=\textwidth]{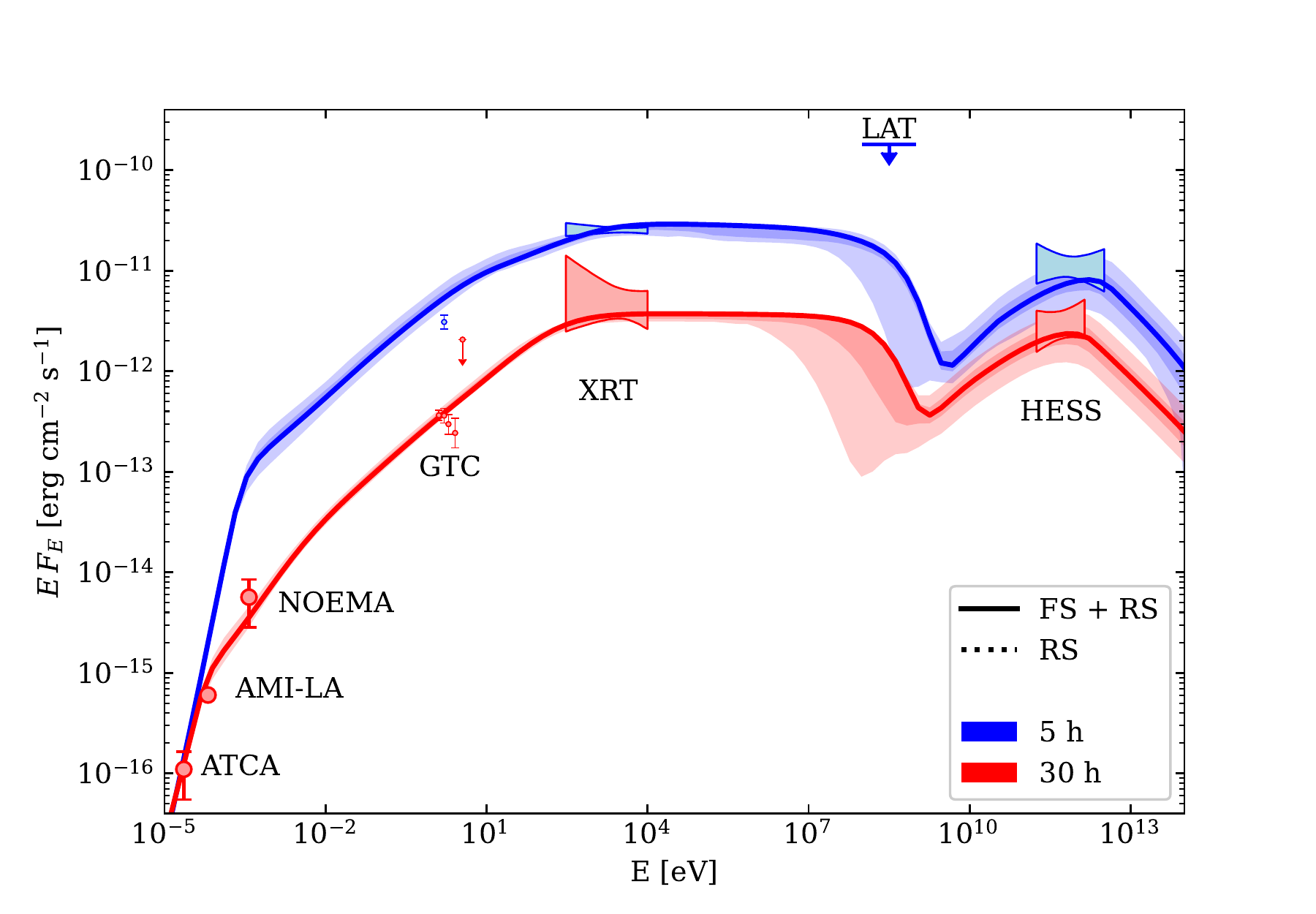}    
\end{center} 
\caption{Predicted spectral energy distributions at the times of the HESS detections. We show with blue (resp.~red) solid lines our model at 5 hours (resp.~30 hours) after the gamma-ray trigger, with 90\% and 50\% credible bands in lighter shades. The HESS 
`butterflies' include the reported \citep{HESS2021} systematic error contribution (summed in quadrature). We also show XRT butterflies at the corresponding times (from our own analysis, see Appendix~\ref{sec:swift_data}), plus GTC optical and NOEMA, ATCA and AMI-LA radio datapoints taken at observing times lying within 0.2 dex.} 
\label{fig:SSC_SED}
\end{figure*}

Figure~\ref{fig:lightcurves} shows the GRB~190829A afterglow light curves in the X-ray, optical and radio bands obtained by combining publicly available data (marked with circles -- see Appendix~\ref{sec:data_from_literature}) with the flux densities measured in our VLBI campaign (marked with stars -- see Appendix~\ref{sec:vlbi_data_reduction}). The solid lines represent the predictions of our best-fit afterglow model (Appendix~\ref{sec:afterglow_model}), where the dashed lines show the contribution from the reverse shock only, while the solid lines also include the forward shock, which dominates the emission at all wavelengths from around one day onwards. In addition, Figure~\ref{fig:SSC_SED} shows the predicted spectral energy distributions at 5 h (blue) and 30 h (red) after the gamma-ray burst, which agree with the emission detected \citep{HESS2019GCN,HESS2021} by the High Energy Stereoscopic System (HESS, butterfly-shaped symbols show one-sigma uncertainties -- including systematics -- when assuming a power-law spectral shape). In our interpretation, therefore, the HESS emission is Synchrotron Self-Compton from the forward shock. Differently from what was reported in the main text of \citet{HESS2021}, we do not find significant photon-photon absorption, at least for our model parameters (see Appendix~\ref{sec:tau_gammagamma_afterglow}). From this modelling, we obtained $\log[(E/n)/\mathrm{erg\,cm^{3}}] = 53.9_{-0.2}^{+0.4}$ (90\% credible level, posterior shown by the red line in Fig.~\ref{fig:E_n_constraint}), in agreement with the VLBI size upper limits, as can also be appreciated from Fig.~\ref{fig:size_evolution}, where the source size evolution entailed by the afterglow emission model (red dashed line) is compared with our source size upper limits. We regard our ability to interpret all the available data self-consistently as a success of the standard gamma-ray burst afterglow model, confirming our general understanding of these sources, but we stress that in order to obtain these results we had to include a number of often overlooked (even though widely agreed upon in most cases) elements in the model.

The results of our afterglow model fitting (see Table~\ref{tab:afterglow_params} and Figure~\ref{fig:mcmc_total}) provided some rather unique insights on the physics of gamma-ray bursts and of the forward and reverse shocks that form as the jet expands into the interstellar medium. Remarkably, we found that the usual simplifying assumption $\chi_\mathrm{e}=1$ in the forward shock is excluded (that is, we were unable to find a statistically acceptable solution when assuming all electrons in the shock downstream to be accelerated to relativistic speeds) and we had $\chi_\mathrm{e}<0.04$ at 90\% credibility when adopting a wide prior $-10<\log(\chi_\mathrm{e})<0$. On the other hand, with such a wide prior we found our uncertainty on the total (collimation-corrected, two-sided) jet kinetic energy to extend towards unrealistically large values $E_\mathrm{jet}=E/(1-\cos\theta_\mathrm{jet})\gtrsim 10^{53}\,\mathrm{erg}$ (assuming two oppositely oriented, identical jets of half-opening angle $\theta_\mathrm{jet}$), corresponding to very small fractions of accelerated electrons $\chi_\mathrm{e}\lesssim 10^{-3}$. When adopting a tighter prior $-2<\log(\chi_\mathrm{e})<0$, motivated by particle-in-cell simulations of relativistic collisionless shocks (which typically find $\chi_\mathrm{e}$ to be around a few per cent, \citealt{Spitkovsky2008,Sironi2011}), we obtained best-fit values consistent within the uncertainties, but the unrealistic-energy tails were removed. In what follows, we report the results for this latter prior choice (we report one-sigma credible intervals unless otherwise stated), while the results for the wider prior are given in the Appendix (Table~\ref{tab:afterglow_params}). 
The jet isotropic-equivalent kinetic energy at the onset of the afterglow is $E = 2.5^{+1.9}_{-1.3}\times 10^{53}\,\mathrm{erg}$ and the jet half-opening angle is $\theta_\mathrm{jet}=15.4_{-0.9}^{+1.3}$ degrees, implying a total jet energy  $E_\mathrm{jet}= 9_{-4}^{+9}\times 10^{51}\,\mathrm{erg}$, which is about one half of the energy in the associated supernova \citep{Hu2020}. Given the observed gamma-ray isotropic equivalent energy $E_\mathrm{\gamma,iso}=(2.91\pm 0.18)\times 10^{50}\,\mathrm{erg}$ (see Appendix~\ref{sec:GBM_reduction}), the implied gamma-ray efficiency is $\eta_\gamma = E_\mathrm{\gamma,iso}/(E_\mathrm{\gamma,iso} + E) = 1.2^{+1.0}_{-0.5}\times 10^{-3}$. This efficiency is much lower than typical estimates for other gamma-ray bursts in the literature \citep{Fan2006,Zhang2007,Wygoda2016,Beniamini2016}, even though we note that a recently published study \citep{Cunningham2020} of GRB~160625B also found a low efficiency when leaving the $\chi_\mathrm{e}$ parameter free to vary. 
The prompt emission efficiency we find is compatible with that expected in the case of internal shocks within the jet \citep{Rees1994} with a moderate Lorentz factor contrast \citep{Kumar1999}. 

The jet bulk Lorentz factor before the onset of the deceleration is $\Gamma_0=57^{+4}_{-5}$. Considering the isotropic-equivalent radiated energy $E_\mathrm{iso}\sim 3\times 10^{50}\,\mathrm{erg}$, this is in agreement with the $\Gamma - E_\mathrm{iso}$ correlation found for long GRBs (see Fig.~\ref{fig:eiso_gamma}, see \citealt{Ghirlanda2018}).

The external medium number density (assumed constant) is relatively low $n=2.1_{-1.0}^{+3.7}\times 10^{-1}\,\mathrm{cm^{-3}}$. This could be tentatively explained by the large offset of the GRB location with respect to the host galaxy centre. Indeed, using the GRB coordinates derived from our VLBI observations and the host galaxy centre position from the 2MASS catalogue \citep{Skrutskie2006}, we measure a separation of 9.6 arcsec, corresponding to a physical projected separation of 14.7 kiloparsec. This is comparable to the largest previously measured offset in long GRBs \citep[][that of GRB~080928]{Blanchard2016}, placing it in principle in the underdense outskirts of its host galaxy. On the other hand, even though the surrounding interstellar medium density may be low, the associated supernova indicates that the progenitor must have been a massive star, which should have polluted the environment with its stellar wind. By contrast, the sharp increase in the flux density preceding the light curve peak as seen in the optical and X-rays is inconsistent with a wind-like external medium, which would result in a much shallower rise \citep{Kobayashi2003}. This places stringent constraints on the properties of the pre-supernova stellar wind, whose termination shock radius \citep{vanMarle2006} must be smaller than the nominal deceleration radius in the progenitor wind $R_\mathrm{dec,w}=E/4\pi A m_\mathrm{p}\Gamma_0^{2} c^2$, where $m_\mathrm{p}$ is the proton mass, $c$ is the speed of light. The parameter $A$ sets the stellar wind density, and can be expressed as a function of the wind mass loss rate $\dot M_\mathrm{w}$ and velocity $v_\mathrm{w}$ as $A=3\times 10^{35} \dot M_\mathrm{w,-5}v_\mathrm{w,3}\equiv 3\times 10^{35} A_\star$, where $\dot M_\mathrm{w,-5}=\dot M_\mathrm{w}/10^{-5}\mathrm{M_\odot/yr}$ and $v_\mathrm{w,3}=v_\mathrm{w}/1000\,\mathrm{km/s}$. Requiring the wind termination shock radius \citep{vanMarle2006}, which depends on the wind properties and also on the external interstellar medium density $n_0$ and on the progenitor lifetime $t_\star$, to be smaller than $R_\mathrm{dec,w}$, we obtain 
\begin{equation}
    \dot M_\mathrm{w,-5} < 3\times 10^{-4} E_\mathrm{52}^{10/13}\Gamma_{0,2}^{-20/13} v_\mathrm{w,3}^{9/13} n_\mathrm{0,2}^{3/13} t_\mathrm{\star,Myr}^{-4/13}
\end{equation}
where $E_\mathrm{52}=E/10^{52}$, $\Gamma_{0,2}=\Gamma_0/100$, $n_{0,2}=n_0/100\,\mathrm{cm^{-3}}$ and $t_{\star,Myr}=t_\star/1\,\mathrm{Myr}$. Inserting our best-fit afterglow parameters, we obtain $\dot M_\mathrm{w,-5} < 7 \times 10^{-2} v_\mathrm{w,3}^{9/13} n_\mathrm{0,2}^{3/13} t_\mathrm{\star,Myr}^{-4/13}$. For the fiducial wind speed, external interstellar medium density (here we set $n_0=100\,\mathrm{cm^{-3}}$ assuming that, despite the large offset, the progenitor was embedded in a star forming region -- but the dependence on this parameter is very weak) and progenitor lifetime parameters, this limits the wind mass loss rate to $\dot M_\mathrm{w}< 7 \times 10^{-7}\,\mathrm{M_\odot\,yr^{-1}}$, which can be achieved only in the case of a very low metallicity, or a low Eddington ratio \citep{Sander2020}. Alternatively, the low wind mass loss rate could be explained as the result of wind anisotropy induced by the fast rotation of the progenitor star \citep{Ignace1996,Eldridge2007}, which would reduce the wind mass loss rate along the stellar rotation axis. 

For the forward shock, we found a relativistic electron power law slope $p_\mathrm{FS}=2.010_{-0.0025}^{+0.0021}$, reminiscent of that expected for first-order Fermi acceleration in non-relativistic strong shocks \citep{Bell1978}, and slightly lower than $p\sim 2.2$ expected for relativistic shocks \citep{Sironi2011}. When $p$ is close to (or below) 2, as in our case, the adopted value of the maximum electron energy $\gamma_\mathrm{max}$ starts impacting the normalisation of the relativistic electron energy spectrum. For this reason, we also fitted an additional free parameter $\gamma_\mathrm{max}/\gamma_\mathrm{min}$, which sets the ratio (assumed constant throughout the evolution) of the maximum to the minimum electron energy in the injected relativistic electron power law. We find $\log(\gamma_\mathrm{max}/\gamma_\mathrm{min})>4.6$ at the 90\% credibile level. The one-sigma credible interval on the fraction of accelerated electrons is $\chi_\mathrm{e,FS}= 2.3_{-1.3}^{+1.1}\times 10^{-2}$ (note that the uncertainty extends down to $\chi_\mathrm{e}\sim 10^{-3}$ when adopting the wider prior -- see Table~\ref{tab:afterglow_params} -- as discussed above). The electron energy density fraction is $\epsilon_\mathrm{e,FS}=3.0^{+2.9}_{-1.7}\times 10^{-2}$, slightly lower than, but comparable to, the expected $\epsilon_\mathrm{e}\sim 0.1$ for mildly relativistic, weakly magnetised shocks \citep{Sironi2011}. On the contrary, the magnetic field energy density fraction is $\epsilon_\mathrm{B,FS}=2.5^{+3.5}_{-1.3}\times 10^{-5}$, in line with previous studies of gamma-ray burst afterglows \citep{BarniolDuran2014,Wang2015}, implying inefficient magnetic field amplification by turbulence behind the shock or a relatively fast decay of such turbulence with the distance from the shock front \citep{Lemoine2013,Lemoine2015}. 
Interestingly, the best-fit values we found for the jet isotropic-equivalent energy $E$, the interstellar medium number density $n$ and the forward shock microphysical parameters $\epsilon_\mathrm{e,FS}$ and $\epsilon_\mathrm{B,FS}$ all closely resemble those found \citep{MAGIC2019IC} for another gamma-ray burst recently detected at TeV energies, GRB~190114C, under the constant external density assumption.

For the reverse shock, we fixed $\chi_\mathrm{e,RS}=1$ as usual to reduce the number of parameters, since we could not constrain it to be lower than this value. We found that, in order to be able to interpret the X-ray and optical peaks at $t\sim 10^{-2}\,\mathrm{days}$ as reverse shock emission (corresponding to the end of reverse shock crossing, see Eq.~\ref{eq:tdec}) without overpredicting \citep[see the typical radio reverse shock light curve shapes in ][which show late-time bumps]{Resmi2016} the later radio data, the magnetic field in the shock downstream must have decayed rapidly after the reverse shock crossed the jet. In particular, we found that the magnetic energy density must have decayed at least as fast as $B^2\propto V^{-\eta_\mathrm{B}}$ with $\eta_\mathrm{B}\geq 6$, where $V$ is the comoving volume of the shell (see Appendix~\ref{sec:afterglow_model} for a detailed description of the assumed dynamics before and after the shock crossing), differently from the usual simplifying assumption \citep{Kobayashi2000lightcurves} that $\epsilon_\mathrm{B}$ remains constant before and after the shock crossing. We consider this reasonable, since the magnetic field is expected \citep{Chang2008} to decay due to Landau damping of the shock-generated turbulence (which produces the magnetic field) after the shock crossing.  For $\eta_\mathrm{B}\geq 6$ our results are independent of the exact value of $\eta_\mathrm{B}$, and we obtained $\epsilon_\mathrm{e,RS}=0.28_{-0.16}^{+0.32}$ and $\epsilon_\mathrm{B,RS}=1.2_{-0.8}^{+1.8}\times 10^{-3}$, and the accelerated electron power law index $p_\mathrm{RS}=2.13_{-0.08}^{+0.04}$.

\subsection{Viewing angle limits}

The inference on the afterglow parameters described so far is based on the assumption of an on-axis viewing angle. On the other hand, a slightly off-axis viewing angle could explain the relatively low luminosity and low peak energy of the observed prompt emission \citep{Sato2021}. This would imply, however, some degree of proper motion in the VLBI images \citep{Fernandez2021}, \textbf{which can be tested in principle with our observations}. Considering the EVN 5 GHz and VLBA 15 GHz epochs only, as these were performed under sufficiently homogeneous observing strategies and shared the same phase-reference calibrator (see Appendix~\ref{sec:vlbi_data_reduction}), the largest displacement compatible at one-sigma with \textbf{the absence of an observed proper motion in our data} is $0.71$ mas (one-sigma upper limit -- including systematic errors -- on the displacement between our first VLBA 15 GHz and our last EVN epoch, which are the two most widely separated in both time and centroid position). At the source distance, this corresponds to $\delta r_\mathrm{max} < 1.088$ parsecs from $t_0=7.89$ to $t_1=69.5$ rest-frame days separating the two observations. \textbf{In order to turn this into a limit on the source properties}, we note that the apparent displacement $\delta r$ of an off-axis jet is bound to be smaller than, or at most equal to, the size increase $\delta s$ of a spherical relativistic blastwave with the same $E/n$ ratio (i.e., the same Sedov length) over the same time, since the jet can be thought of as a portion of that sphere. Again using the self-similar expansion law from \citet{Granot1999}, a relativistic blastwave would need to have  $\log[(E/n)/\mathrm{erg\,cm^{3}}] \geq 59.3$ in order to produce an expansion $\delta s \geq \delta r_\mathrm{max}$ over the same time range, which is well beyond any conceivable value for a gamma-ray burst. This means that our astrometric measurements are not sufficiently precise to exclude any viewing angle. 

\begin{figure}[t]
\includegraphics[width=\columnwidth]{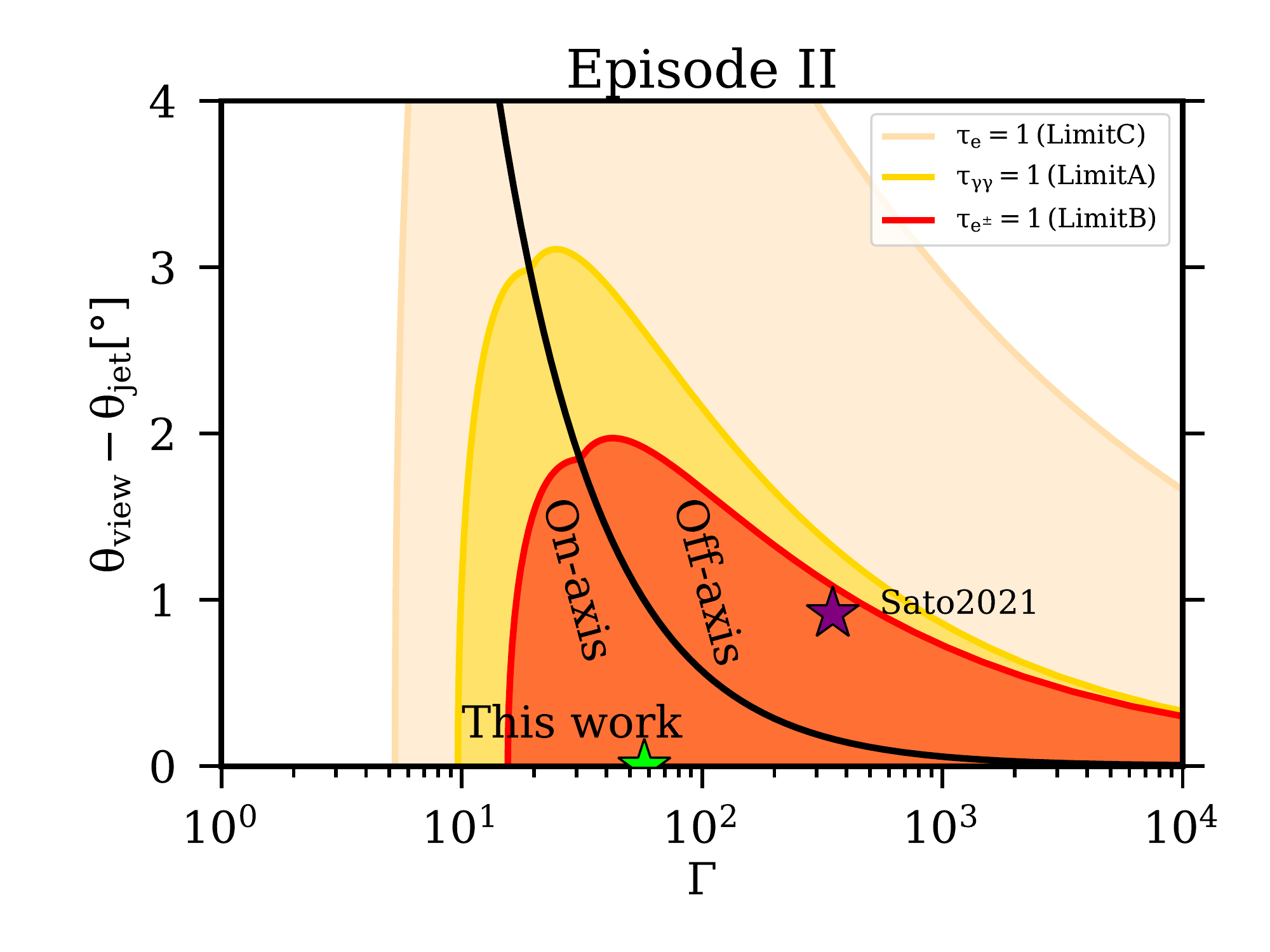}
\caption{Limit on the viewing angle from compactness arguments on the prompt emission. Shaded areas show the allowed regions on the ($\Gamma$, $(\theta_{\rm view}-\theta_{\rm jet})$) plane, derived by the compactness limits A (photon-photon pair-production), B (scattering off $e^{\pm}$), and C (scattering off $e^-$ associated to baryons). The solid black line ($\Gamma \theta = 1$) separates on-axis observers from off-axis ones. The green star marks the bulk Lorentz factor $\Gamma$ value inferred from the afterglow lightcurve modelling, while the purple star marks the parameters used in \citet{Sato2021}. Both solutions are in the allowed region that ensures the source to be transparent to the observed high-energy prompt emission photons.}
\label{fig:compactness}
\end{figure}

A relatively tight limit on the viewing angle can be obtained, on the other hand, by requiring the jet to be optically thin to the photons we observed during the prompt emission. In particular, we performed the calculation of the optical depth to $\gamma$-ray photons for an arbitrary viewing angle and jet Lorentz factor \citep{Matsumoto2019b}, given the observed spectrum. We focused on the brightest emission episode, namely episode II, that provides the most stringent limit on the viewing angle.
Photons of energy $E$ must have been able to escape from the emitting region and not pair-annihilate with other photons of energy $\geq (\delta m_e c^2)^2 / E$, where $\delta$ is the relativistic Doppler factor $\delta = [\Gamma(1-\beta \cos\theta)]^{-1}$ (limit A); they must not have been scattered off by pairs produced by the annihilation of other high energy photons (limit B); and they must not have been scattered off by the electrons associated with the baryons in the outflow (limit C). 
The first two sources of opacity depend on the observed spectrum, while the third one depends on the matter content of the jet, which we conservatively assumed to be the lowest compatible with the observed spectrum.
Given the prompt emission spectrum observed in episode II, we computed the optical depth as a function of the bulk Lorentz factor $\Gamma$ and the viewing angle $\theta_{\rm view}$ for limits A, B and C following \citet{Matsumoto2019b} and assumed an emission duration $\delta_{t} = 3 s$, which corresponds to the brightest peak in the emission episode. Figure~\ref{fig:compactness} shows the regions on the $\Gamma$, $\theta_{\rm view} - \theta_{\rm jet}$ for which the optical depths are smaller than unity for the three limits. The solid black line corresponds to $\Gamma (\theta_\mathrm{view}-\theta_\mathrm{jet})= 1$, therefore dividing the plot into on- and off-axis regions (inside or outside the relativistic beaming cone of material within the jet border). As shown in the plot, the value of $\Gamma$ derived from our afterglow modelling (represented by the green star) is within the relatively small allowed region. The resulting upper limit on the viewing angle is $\theta_\mathrm{view}-\theta_\mathrm{jet}\lesssim 2^{\circ}$.
Adopting the jet opening angle $\theta_{\rm jet} = 15^{\circ}$ obtained from the afterglow modelling, a viewing angle greater than 17$^{\circ}$ would not be compatible with the observed emission.

Recently, a two-component jet model has been proposed \citep{Sato2021} to explain the multi-wavelength observations of GRB~190829A. In particular, a narrow ($\theta_{\rm jet} = 0.015 \, \mathrm{rad} = 0.86^{\circ}$) and fast ($\Gamma = 350$) jet was used to reproduce the bumps observed in the optical and X-rays at $t \sim 1.4 \times 10^{3} \,\mathrm{s}$ from the trigger time, while a wide ($\theta_{\rm jet} = 0.1 \, \mathrm{rad} = 5.73^{\circ}$) and slow ($\Gamma = 20$) co-axial jet should explain the late ($t \gtrsim 10^{5} \,\mathrm{s}$) X-ray and radio emission. In this scenario, the observer is at an angle $\theta_{\rm view} =  0.031 \, \mathrm{rad} \, (1.78^{\circ})$ with respect to the jet axis. Since the authors of that work point out that the narrow jet could be responsible for the prompt emission of both the episodes I and II, we also applied the compactness argument to this solution for comparison. As shown in Figure~\ref{fig:compactness}, the parameters they assumed for the narrow jet are still inside the allowed region, although quite close to its limit, and therefore the solution with an off-axis narrow jet as the source of the observed gamma-rays is not ruled out from the compactness argument.
\section{Summary and discussion}

Our VLBI observations and analysis provide evidence in support of the GRB~190829A afterglow being produced by a relativistic blastwave, at least at $t\geq 9\,\mathrm{days}$. We found that a forward plus reverse shock afterglow model, assuming an on-axis viewing angle and a uniform external medium density, is able to reproduce the observed light curves from the gamma rays down to the radio at 1.4 GHz, provided that only a relatively small fraction $\chi_\mathrm{e}\lesssim \mathrm{few}\times 10^{-2}$ of the electrons have been accelerated to relativistic speeds in the forward shock, and that the magnetic field in the reverse-shocked jet decays rapidly after the shock crossing. The required external medium density is relatively low, which points to a very weak progenitor stellar wind. The size evolution entailed by the model is in agreement with the limits set by our VLBI observations. On the other hand, while our calculations are based on the assumption of an on-axis jet, our analysis cannot exclude a viewing angle slightly off the jet border, in which case our derived parameters (especially those related to the reverse shock) would possibly require some modification. The jet and forward shock parameters  obtained from our analysis are similar to those found \citep{MAGIC2019IC} for GRB~190114C in the constant external density scenario. 

As a final note, we point out that other interpretations of this GRB, differing from the one presented in this paper, have been proposed in the literature. The main point of qualitative disagreement among these interpretations is the X-ray/optical peak at $t\sim 10^{-2}\,\mathrm{days}$ (i.e.~around $10^3$ s): \cite{ZhangT2020} attribute it to late central engine activity; \cite{ZhangL2021} invoke the interaction of the blastwave with a pre-accelerated, electron-positron-pair enriched shell formed due to annihilation of prompt emission photons, partly scattered by the dusty external medium; \cite{Fraija2020} propose instead a magnetar spin-down-powered origin. Given that the reverse shock is a natural consequence of the jet interaction with the external medium, our interpretation (within which we are able to explain all the data self-consistently) can be preferred with respect to these based on Occam's razor. Finally, \cite{Rhodes2020} proposed a forward plus reverse shock interpretation. In contrast with us, though, they attribute the 15.5 GHz data at $t>1\,\mathrm{day}$ to a reverse shock in the thick shell regime. We note that, in this regime, the reverse shock emission would peak at the end of the prompt emission (around 70 s post-trigger), so that the X-ray/optical peak would remain unexplained.

\section{Acknowledgements} OS thanks Marco Landoni and the Information and Communication Technologies (ICT) office of the Italian National Institute for Astrophysics (INAF) for giving access to the computational resources needed to complete this work; he also acknowledges the Italian Ministry of University and Research (MUR) grant `FIGARO' (1.05.06.13) and the INAF-Prin 2017
(1.05.01.88.06) for financial support.
 TA, PM and YKZ made use of the computing resources of the China SKA Regional Centre prototype under support from National Key R\&D Programme of China (grant number 2018YFA0404603), NSFC (12041301) and Youth Innovation Promotion Association of CAS.  
 GG acknowledges support from MIUR, PRIN 2017 (grant 20179ZF5KS)
 BM acknowledges support from the Spanish Ministerio de Econom\'ia y Competitividad (MINECO) under grant AYA2016-76012-C3-1-P and from the Spanish Ministerio de Ciencia e Innovaci\'on under grants PID2019-105510GB-C31 and CEX2019-000918-M of ICCUB (Unidad de Excelencia ``Mar\'ia de Maeztu'' 2020-2023). 
The European VLBI Network (EVN) is a joint facility of independent European, African, Asian, and North American radio astronomy institutes. Scientific results from data presented in this publication are derived from the following EVN project code(s): EG010. 
e-VLBI research infrastructure in Europe is supported by the European Union’s Seventh Framework Programme (FP7/2007-2013) under grant agreement number RI-261525 NEXPReS.
e-MERLIN is a National Facility operated by the University of Manchester at Jodrell Bank Observatory on behalf of STFC.
The research leading to these results has received funding from the European Commission Horizon 2020 Research and Innovation Programme under grant agreement No. 730562 (RadioNet).
The National Radio Astronomy Observatory is a facility of the National Science Foundation operated under cooperative agreement by Associated Universities, Inc. Scientific results from data presented in this publication are derived from the following VLBA project codes: BA140, BO062.
This work made use of the Swinburne University of Technology software correlator, developed as part of the Australian Major National Research Facilities Programme and operated under licence.
This work made use of data supplied by the UK Swift Science Data Centre at the University of Leicester.

\paragraph{Data availability.} The European Very-Long Baseline Interferometry Network data (PIs: Ghirlanda \& An) whose analysis has been presented in this study are publicly available at the EVN Data Archive at JIVE \footnote{\url{http://www.jive.nl/select-experiment}} under the identifiers RG010A, RG010B and RG010C. The Very Long Baseline Array data are publicly available at the National Radio Astronomy Observatory (NRAO) Data Archive \footnote{\url{https://science.nrao.edu/facilities/vlba/data-archive}} under the identifiers BO062 (5 GHz epochs, PI: Orienti) and BA140 (15 GHz epochs, PI: An). The Neil Gehrels Swift Observatory data analysed in this study is publicly available at the UK Swift Data Centre \footnote{\url{https://www.swift.ac.uk/xrt\_spectra}} at the University of Leicester. The Fermi/GBM data analysed in this study are publicly available at the National Aeronautics and Space Administration (NASA) High-Energy Astrophysics Science Archive Research Centre (HEASARC) Fermi GBM Burst Catalog \footnote{\url{https://heasarc.gsfc.nasa.gov/W3Browse/fermi/fermigbrst.html}}. All reduced data and computer code are available from the corresponding authors upon reasonable request.

\appendix

\section{VLBI observations and data analysis}

\subsection{VLBA and EVN observations and data reduction}\label{sec:vlbi_data_reduction}

We performed rapid-response VLBI observations of GRB~190829A with the Very Long Baseline Array (VLBA) and with the European VLBI Network (EVN) plus the enhanced Multi-Element Remotely Linked Interferometry Network (e-MERLIN). All the observations were carried out in phase-referencing \citep{Beasley1995} mode.

\newcommand{\grb}{GRB~190829A}
\newcommand{\calone}{J0257--1212}
\newcommand{\caltwo}{J0300--0846}

\begin{table*}[t]
\caption{\textbf{Summary of VLBI observations and image parameters.}} \label{tab:obs}
\centering 
\small
\begin{tabular}{ccccc}
\hline\hline
UT (duration)       &  Freq. & VLBI Network$^a$   &  Synth.~beam                      & RMS: $\sigma_{\rm i}$ \\ 
~                   &  (GHz) &                    & (Major~$\times$~Minor, PA)        & ($\mu$Jy~beam$^{-1}$) \\
\hline
Sep 17, 22:30 (8~h) &  4.99  & EVN$+$e-MERLIN     &  3.38~$\times$~2.16~mas$^2$, $+$35$^\circ$.7 & 15.3  \\
Oct 15, 21:00 (8~h) &  4.99  & EVN$+$e-MERLIN     &  3.66~$\times$~2.56~mas$^2$, $+$23$^\circ$.2 & 10.4  \\
Nov 12, 19:00 (8~h) &  4.99  & EVN$+$e-MERLIN     &  4.16~$\times$~2.79~mas$^2$, $+$19$^\circ$.6 & 11.4  \\
\hline
Sep 07, 07:56 (6~h) & 15.39  & VLBA               &  1.76~$\times$~0.60~mas$^2$, $-$10$^\circ$.8 & 44.4  \\
Oct 03, 06:14 (6~h) & 15.17  & VLBA               &  1.86~$\times$~0.68~mas$^2$, $-$9$^\circ$.0  & 31.5  \\
\hline
Sep 11, 07:30 (6~h) &  4.98  & VLBA               &  2.16~$\times$~2.16~mas$^2$,  $+$0$^\circ$.0 & 33.8  \\
Oct 16, 05:30 (6~h) &  4.98  & VLBA               &  3.32~$\times$~1.31~mas$^2$,  $-$1$^\circ$.0 & 17.1  \\
Nov 17, 03:15 (6~h) &  4.88  & VLBA               &  5.44~$\times$~2.02~mas$^2$,  $-$4$^\circ$.5 &  9.8  \\
Dec 24, 00:45 (6~h) &  4.88  & VLBA               &  3.33~$\times$~1.26~mas$^2$,  $-$2$^\circ$.5 & 12.6  \\
\hline
\end{tabular}

$^a$The full list of VLBI stations is provided in the text. 
\end{table*}

\label{sec_evn_obs}

EVN plus e-MERLIN observations at 5~GHz were performed under project code RG010 (PI: Ghirlanda~G. \& An~T.) in three epochs (September 17, October 15 and November 12, 2019), with a total of 20 participating telescopes, namely Jodrell Bank MK II (\texttt{Jb}), Westerbork single antenna (\texttt{Wb}), Effelsberg (\texttt{Ef}), Medicina (\texttt{Mc}), Onsala (\texttt{On}), Tianma (\texttt{T6}), Toru\'n (\texttt{Tr}), Yebes (\texttt{Ys}), Hartebeesthoek (\texttt{Hh}), Svetloe (\texttt{Sv}), Zelenchukskaya (\texttt{Zc}), Badary (\texttt{Bd}), Irebene 16 m (\texttt{Ib}), Irebene 32 m (\texttt{Ir}), Cambridge (\texttt{Cm}), Darnhall (\texttt{Da}), Pickmere (\texttt{Pi}),  Defford (\texttt{De}), Knockin (\texttt{Kn}) and Kunming (\texttt{Km}). Stations that missed the observations were \texttt{T6, Ys, Kn} in the first epoch, \texttt{Tr, Km} in the second epoch and \texttt{Sv, Km} in the last epoch. The EVN observations were carried out in electronic-VLBI (e-VLBI) mode \citep{Szomoru2008}, and the data correlation was done in real time by the EVN software correlator (\textsc{SFXC}, \citealt{Keimpema2015}) at the Joint Institute for VLBI ERIC (JIVE) using an integration time of 1~s and a frequency resolution of 0.5~MHz. The results are summarised in Table~\ref{tab:obs}. 

In the first epoch, we observed two phase calibrators \calone{} and \caltwo{}. \calone{} had a correlation amplitude \citep{Charlot2020} of $\gtrsim$0.2~Jy at 8.4~GHz  and an angular separation of 3$^\circ$.24 away from the target source on the plane of the sky. \caltwo{} had a correlation amplitude of $\sim$0.02~Jy at 5~GHz on the long baselines \citep{Petrov2020} and a separation of 0$^\circ$.56. The cycle times for the nodding observations of \calone{} and GRB~190829A were about three minutes at the lower observing elevation in the first and last two hours and about six minutes at the higher elevation in the middle four hours. The secondary calibrator \caltwo{} was observed for a short 2-minute scan every three cycles. In our observing strategy, the nearby weak calibrator \caltwo{} was the phase-referencing calibrator, and the bright calibrator \calone{} was mainly used to significantly boost the phase coherence time to about one hour in the post data reduction. 

Because the closer calibrator \caltwo{} also had high correlation amplitude on the long baselines in the first-epoch observation, we optimised the observing strategy in the remaining two epochs: \caltwo{} was observed more frequently as a traditional phase-referencing calibrator, and \calone{} was observed as a fringe finder for only a few short scans. The cycle times were increased to about four minutes at lower elevations and about seven minutes at higher elevations.   

The data were calibrated with the National Radio Astronomy Observatory (NRAO) software package Astronomical Image Processing System (\textsc{AIPS}, \citealt{Greisen2003}). We first flagged out some off-source or very low-sensitivity visibility data. In the first epoch, the e-MERLIN stations \texttt{Cm} and \texttt{De} had unusually high fringe rates ($\gtrsim$10~mHz) owing to variable delays of their optical cables. Those data were excluded to avoid poor phase connections and some low-level baseline-based errors. \textit{A-priori} amplitude calibration was done with properly smoothed antenna monitoring data (system temperatures and gain curves) or nominal system equivalent flux densities. The ionospheric dispersive delays were corrected by using the maps of total electron content provided by the Global Positioning System satellite observations. The time-dependent phase offsets due to the antenna parallactic angle variations were removed. We aligned the phases across the sub-bands via iterative fringe-fitting with a short scan of the calibrator data. After phase alignment, we combined all the sub-band data in the Stokes $RR$ and $LL$, then ran the fringe-fitting with a sensitive station as the reference station and applied the solutions to all the related sources. In the first epoch, after transferring the fringe-fitting solutions from \calone{} to both \caltwo{} and GRB~190829A, we also ran fringe-fitting on \caltwo{} to solve for only phases and group delays, and then transferred the solutions to GRB~190829A. In this additional iteration, we found that \texttt{Tr} data had poor phase connections in the last four hours and \texttt{Ib} data had large residual delays ($\sim$1~ns) probably due to the uncertainty of their antenna positions or poor weather condition during the observation. Because of these issues, we excluded these problematic data. Finally, bandpass calibration was performed. All the above calibration steps were scripted in the \textsc{ParselTongue} \citep{Kettenis2006} interface. 

We imaged the calibrators \calone{} and \caltwo{} through iterative model fitting with a group of delta functions (point source models), and self-calibration in \textsc{Difmap} \citep{Shepherd1994}. With the input source images, the fringe-fitting and the self-calibration were re-performed in \textsc{AIPS} via a \textsc{ParseTongue} script. All these phase and amplitude solutions were also transferred to the target source data by linear interpolation. The final imaging results of the calibrator \caltwo{} are \textbf{shown in supplementary figures available in the associated Zenodo repository \citep{zenodo_supplementary}}. They show a one-sided core-jet structure toward the north. The total flux densities are $\sim$34~mJy on September 17, $\sim$41~mJy on October 15, and $\sim$37~mJy on November 12. The compact radio core was modelled by a single point source. In the phase-referencing astrometry, we used the radio peak as the reference point, $03^{\rm h}00^{\rm m}19^{\rm s}.5876$, $-08^\circ46'10''.174$ (J2000). Compared to the latest VLBI global solutions in the radio fundamental catalogue (RFC,  \citealt{Petrov2020}) 2020b, the correction is quite small ($\Delta$RA~=~$+0^{\rm s}.000012\pm0.00003$, $\Delta$Dec~=~$+0''.00053\pm0''.00110$) and thus dropped out in the differential astrometry. The bright calibrator \calone{} had flux densities $\sim$320~mJy on September 17, $\sim$360~mJy on October 15, $\sim$320~mJy on November 12.  

We imaged GRB~190829A in \textsc{Difmap} without self-calibration. To avoid bandwidth-smearing effects, we shifted the target to a position close ($<$1~mas) to the image centre with the \texttt{AIPS} task \texttt{UVFIX}. To improve the phase coherence, we excluded the data observed at low elevations, i.e.  $\leq$15$^{\circ}$.  

We also carried out VLBA observations of \grb{} (project code: BA140, PI: An, T.) at 15~GHz in two epochs (September 7 and October 3, 2019, 6 hr each). All the ten VLBA antennas were used during the observations, namely Hancock (\texttt{Hn}), North Liberty (\texttt{Nl}), Fort Davis (\texttt{Fd}), Los Alamos (\texttt{La}), Pie Town (\texttt{Pt}), Kitt Peak (\texttt{Kp}), Owens Valley (\texttt{Ov}) and Brewster (\texttt{Br}). The data were correlated by a distributed FX-style software correlator (\texttt{DiFX}, \citealt{Deller2007}) at the National Radio Astronomy Observatory. The output data had an integration time of 1~s and a frequency resolution of 0.5~MHz. Table~\ref{tab:obs} summarises the results of these observations.

The VLBA 15-GHz observations of \grb{} had the same observing strategy as the first-epoch EVN plus e-MERLIN observations. Both \calone{} and \caltwo{} were observed. At 15~GHz, we used two cycle times: $\sim$110~s (scan lengths: 20~s for \calone{}, 70~s for \grb{} or \caltwo{}) in the first and last hour, and $\sim$140~s (scan lengths: 20~s for \calone{}, 100~s for \grb{} or \caltwo{}) in the middle four hours. \caltwo{} was treated as a pseudo target during the observations. Every six or seven cycles, there was a cycle for \caltwo{}. The bright ($>1$~Jy at 15~GHz) radio sources 0234$+$285 and NRAO~150 were observed as the fringe finders. While similar to the EVN strategy, the procedure worked overall better because of the shorter cycle times and the higher mean elevation of the Dec.\ $\sim -9^\circ$ target for the VLBA stations.

Post data reduction was carried out with \textsc{AIPS} and \textsc{Difmap} installed in the China SKA Regional Centre prototype \citep{An2019}. The calibration strategy in \textsc{AIPS} was basically the same as that used in the EVN plus e-MERLIN observations described above. The correlator digital correction was applied when the data were loaded into \textsc{AIPS}. Deviations in cross-correlation amplitudes owing to errors in sampler thresholds were corrected using the auto-correlation data. The atmospheric opacity was solved and removed using the system temperature data measured at each station. \textit{A-priori} amplitude calibration was made with properly smoothed system temperatures and gain curves. The correction on the Earth orientation parameters was applied. Ionospheric dispersive delays were corrected according to maps of total electron content. Phase offsets due to antenna parallactic angle variations were removed. After the fringe-fitting on the fringe finders, bandpass calibration was performed. We ran a global fringe-fitting on \calone{} and applied the solutions to both \caltwo{} and \grb{}. After that, we ran another global fringe fitting on the weak calibrator \caltwo{}, by switching off the solutions of the residual fringe rate and used a low signal-to-noise ratio cutoff of 3. With such setups, we got more accurate phase and delay solutions, in particular for the long-baseline data.  

We then imaged the calibrators \calone{} and \caltwo{} in \textsc{Difmap}. The self-calibration and imaging procedure at 15~GHz was the same as that at 5~GHz. \calone{} shows a one-sided core-jet structure with total flux densities of 0.36~$\pm$~0.02~Jy in the first epoch and 0.38~$\pm$~0.02~Jy in the second one. The correlation amplitude is quite high, $\gtrsim$0.15~Jy on all the baselines. 

The imaging results of \caltwo{} are \textbf{displayed in supplementary figures available in the associated Zenodo repository \citep{zenodo_supplementary}}. \caltwo{} has a core-jet structure with total flux densities $\sim$22~mJy in the first epoch and $\sim$26~mJy in the second epoch. In both epochs, the visibility data of \caltwo{} could be simply fitted with four point-source models. After the two calibrator images were made, we re-ran fringe-fitting to remove source structure-dependent phase errors. To improve the amplitude calibration further, we also applied the self-calibration amplitude solutions of \calone{} to the data of \grb{} in \textsc{AIPS}. 

\begin{figure}[t]
\centering
\includegraphics[width=0.6\textwidth]{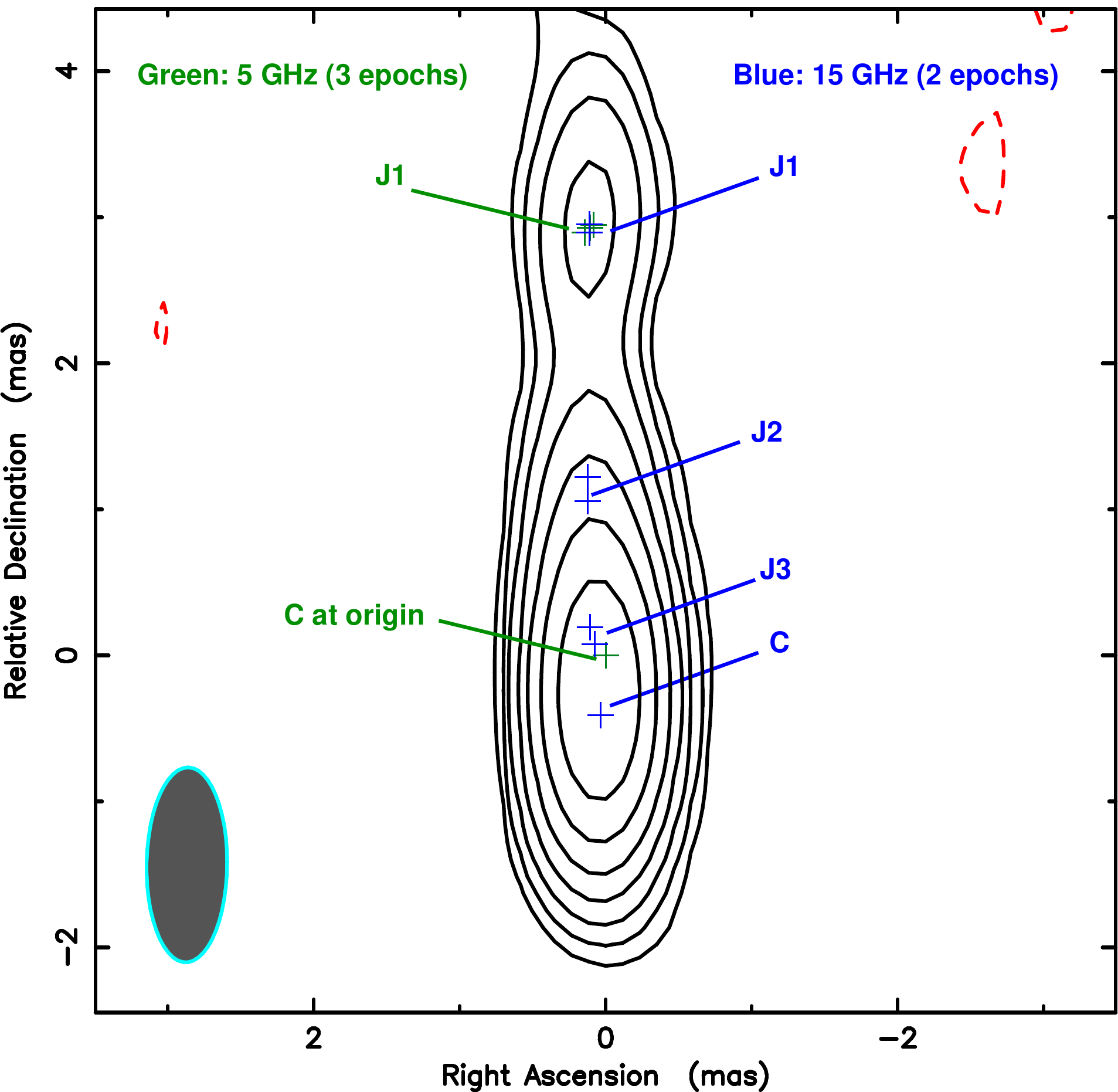} \\

\caption{The frequency-dependent shift of the initial reference point C from 5 GHz to 15 GHz. The best-fit point-source models are plotted as green crosses at 5~GHz and blue crosses at 15~GHz. The contours show the 15-GHz image at the second epoch. Using the mean position of the optically thin, discrete and compact jet component J1 as the reference point, the shift was derived and corrected in the joint astrometry. }

\label{fig:c2k}
\end{figure}

As a starting point, we used the radio core of \caltwo{} as the reference position at 15~GHz as we did for the 5~GHz. We noticed, though, that the partially self-absorbed radio core of \caltwo{} has a frequency-dependent positional shift (the so-called 'core shift' effect, \citealt{Kovalev2008}) mainly along the jet direction (see Fig.~\ref{fig:c2k}). Using the mean position of the compact (size $\leq$0.22~mas), relatively discrete and steep-spectrum component J1 as the reference point, we corrected the frequency-dependent shift of the initial reference point C from 5 GHz to 15 GHz. The 15-GHz component C has a mean positional shift of $\Delta$RA~=~0.035~$\pm$~0.008~mas, $\Delta$Dec~=~$-$0.408~$\pm$~0.019~mas with respect to the 5-GHz component C. The jet component J1 had flux densities $5.3\pm0.3$~mJy at 5~GHz and $1.6\pm0.1$~mJy at 15~GHz, implying an optically thin spectrum ($dF/d\nu\propto \nu^{-1.04\pm0.07}$) and its position has therefore negligible frequency dependence. Because of the high redshift \citep{Hewett2010} $z=1.862$ and the short time baseline, the positional shift of J1 between any two epochs is quite small, $\leq0.063$~mas in RA or Dec.

We imaged \grb{} at 15~GHz adopting the same procedure as for the EVN data. To avoid bandwidth smearing effects, we also shifted its position before doing any average. To improve the phase-referencing precision, we excluded the data observed at the low elevations of $\leq$50$^\circ$ for the near-sea station \texttt{SC} and $\leq$30$^\circ$ for the rest stations. 

An additional set of VLBA observations of GRB\,190829A triggered by the TeV detection \citep{HESS2019GCN} were
carried out at 5 GHz (C-band; project code BO062, PI: Orienti, M.) for a total of four epochs between September and December 2019 (see Table~\ref{tab:obs} for
details). Observations were performed with a recording bandwidth of 128 MHz and a 2048 Mbps data rate, with the exception of the last two epochs which made use of 4 Gbps data rate. We centred the observation of the first epoch at \citep{GCN25584} RA (J2000) = 02$^h$58$^m$10$^s$.510, and Dec (J2000) = $-$08$^\circ$57$'$28$''$.44, whereas the observations
of the following epochs were centred at RA (J2000) = 02$^h$58$^m$10$^s$.5219 and Dec (J2000) = $-$08$^\circ$57$'$28$''$.0933 based on the results of the first observation.

During each observing run, the target source was observed for about 4 hr in phase-referencing mode. Scans on the target were
bracketed by scans on the phase-calibrator J0257$-$1212. In addition, every hour we spent a 3-min scan on the phase-referencing check source J0253$-$1200 at about 1$^\circ$.07 from the
phase-reference source. Considering the time on the target source,
calibrations and overhead, the total observing time for each run was about 6 hr.

Editing and \textit{a priori} calibration was performed following standard procedures as described above and also in the \textsc{AIPS} cookbook, correcting for ionospheric dispersive delays, digital sampling corrections, parallactic angle variations, instrumental delays. We calibrated the bandpass using a scan on 3C\,84 in which all the antennas had
good data.
Amplitudes were calibrated using the antenna system temperatures and antenna gains. Uncertainties on the amplitude scale, $\sigma_{\rm cal}$, were about 7 per cent. We performed
global fringe fitting to correct for residual fringe delays and
rates. Since the target source is too faint for fringe-fitting, we applied the solutions of the phase-reference calibrator J0257$-$1212 to the target and the check source J0253$-$1200. We also fringe-fitted the check source in order to compare the flux density obtained with and without fringe-fitting. The two values were in good agreement.

Images were produced using the task \texttt{IMAGR} in \textsc{AIPS}. The source is clearly detected in all epochs. 
We performed the analysis for determining the astrometry, but the angular separation to the phase referencing calibrator was proven to be too large, preventing an accurate determination of the position
(uncertainties of about 0.3 mas), therefore these data were not constraining for what concerns any potential source projected motion.

The final imaging results of \grb{} are shown in \textbf{supplementary figures available in the associated Zenodo repository \citep{zenodo_supplementary}}. The synthesised beams and the image noise levels are reported in Table~\ref{tab:obs}. The target \grb{} was clearly detected in all the nine epochs with signal-to-noise ratios (SNR) ranging from 10 to 31. 
Moreover, the VLBI observations at the same observing frequency had quite similar coverages of the ($u, v$)-plane.  

The peak flux densities and the circular Gaussian model fitting results are tabulated in Table~\ref{tab:fit}. 

Besides the fitting uncertainties reported in Table~\ref{tab:fit}, we included in the error budget systematic positional uncertainties of 0.051~mas in RA and 0.075~mas in Dec for the EVN and VLBA 15~GHz epochs, and 0.3~mas in RA and 0.4~mas in Dec for the VLBA 5~GHz epochs. These stem from a statistical study \citep{Paragi2013} of four-epoch VLBA phase-referencing observations of a pair of extra-galactic sources (J1707$-$1415 and NVSS3, separation: 1$^\circ$.89) at 5~GHz, whose reported $1\sigma$ scatters are 0.17~mas in RA and 0.25~mas in Dec. Our systematic uncertainty estimates were derived by re-scaling these values by the ratio (0.3 for EVN and VLBA 15 GHz; 1.7 for VLBA 5 GHz) of our target source--phase reference source angular separation to that of the cited study, due to the fact that systematic positional uncertainties are generally proportional to angular separations in VLBI phase-referencing astrometry \citep{Franz2015}. Because \grb{} has a relatively low Declination and there are more East-West long-baseline data, as shown in \textbf{supplementary figures available in the associated Zenodo repository \citep{zenodo_supplementary}}, the astrometry precision in RA is always better than that in Dec. Compared to the EVN astrometry at 5~GHz, the VLBA astrometry at 15~GHz might have somewhat smaller systematic errors because of the higher observing elevation at most VLBA stations and the more uniform antenna sensitivities.

As a side note, we have searched for compact radio components in the central region ($2.46 \times 2.46$~arsec$^2$) of the host galaxy \citep{Heintz2019,Rhodes2020} SDSS~J025810.28$-$085719.2 with the wide field imaging function provided by the {\sc AIPS} task {\tt IMAGR}. We find no compact radio emission with a brightness $\ge 0.056$~mJy\,beam$^{-1}$ ($6\sigma$) at 5~GHz. To search for any extended radio emission, we also tried to use a taper of 0.3 at a ($u$, $v$) radius of 5 mega-wavelengths. With a large beam size of $30 \times 22$ mas$^{2}$, still no emission above $5\sigma$ ($\ge 0.14$~mJy\,beam$^{-1}$) was seen in the dirty maps.   

\subsection{VLBI data source model fitting}\label{sec:vlbi_source_fit_method}
In order to obtain detailed information about the source total flux density, size and position from each of our VLBI epochs, we fitted the calibrated visibility data adopting a Markov Chain Monte Carlo (MCMC) approach. We adopted a simple Gaussian likelihood model, namely
\begin{equation}
\begin{array}{l}
    \ln\mathcal{L}(x) = -\frac{1}{2} \sum_{i=0}^{N}w_i\left[\left(\mathcal{V}_{R,\mathrm{m}}(u_i,v_i,x)-\mathcal{V}_{R,i}\right)^2\right. +  \\
    +\left.\left(\mathcal{V}_{I,\mathrm{m}}(u_i,v_i,x)-\mathcal{V}_{I,i}\right)^2\right] 
\end{array}
\end{equation}
where $\mathcal{V}_{R,i}$ and $\mathcal{V}_{I,i}$ are the real and imaginary part, respectively, of the $i$-th visibility measurement, corresponding to position $(u_i,v_i)$ on the $(u,v)$ plane, and $w_i$ is its \textsc{AIPS}-determined data weight (corresponding to the reciprocal of the square of the associated uncertainty). $\mathcal{V}_{R,\mathrm{m}}(u,v,x)$ and $\mathcal{V}_{I,\mathrm{m}}(u,v,x)$ are the real and imaginary parts of the model source visibility, which we took as a circular Gaussian, evaluated at point $(u,v)$ with parameters $x=(f_\nu,s,\rho,\delta)$, $f_\nu$ being the total flux density at the observing frequency, $s$ the full width at half maximum (FWHM), and $\rho$ and $\delta$ the spherical offsets of the source with respect to the phase centre. With these definitions, one has
\begin{equation}
    \mathcal{V}_\mathrm{m} = f_\nu e^{-2\pi^2 \left(\frac{s}{\sqrt{8\ln 2}}\right)^2\left(u^2+v^2\right)-2\pi i (u\rho + v\delta)}
\end{equation}
where $i=\sqrt{-1}$. We sampled the posterior probability of the parameters using the
\textsc{emcee} \citep{Foreman-Mackey2013} python package and adopting a uniform prior on all parameters, with the constraints $f_\nu>0$ and $10^{-6}<s/\mathrm{mas}<10$. We initialised \textsc{emcee} with the best-fit parameters obtained by fitting the source to a circular Gaussian in \textsc{difmap}, and run $10^4$ iterations of the MCMC with 8 walkers, for a total of $8\times 10^{4}$ evaluations of the posterior probability density, of which we discard the initial half as burn in. Corner plots constructed using the resulting posterior samples are \textbf{available on Zenodo \citep{zenodo_supplementary}}. We took the parameter values corresponding to the sample with the highest posterior probability density as our best fit, we estimated the one-sigma credible range of each parameter as the smallest interval containing 68\% of the marginalised posterior probability, and the 95\% credible size upper limits as the 95-th percentile of the posterior samples. All results are reported in Table~\ref{tab:fit}, and the size upper limits are shown in Fig.~\ref{fig:size_evolution}.

\input{Imaging-table}

\subsection{VLBI source parameter estimation: validation on simulated sources}\label{sec:vlbi_source_fit_validation}
In order to validate our Bayesian parameter estimation approach, we ran our MCMC fitting procedure on several simulated datasets to check whether (and how well) the injected source parameters were recovered. The simulated observations were created by adding a fake circular Gaussian source to the calibrated visibilities of our October 03 VLBA 15 GHz observations (the choice of this particular observation was based on its low SNR, which ensured a minimal interference of the actual GRB source on our results). We performed the experiment several times, varying the SNR between 15 and 120 and the size of the fake source between 0.1 and 3 times the synthesised beam size. 
\begin{figure}[t]
 \centering
 \includegraphics[width=0.6\textwidth]{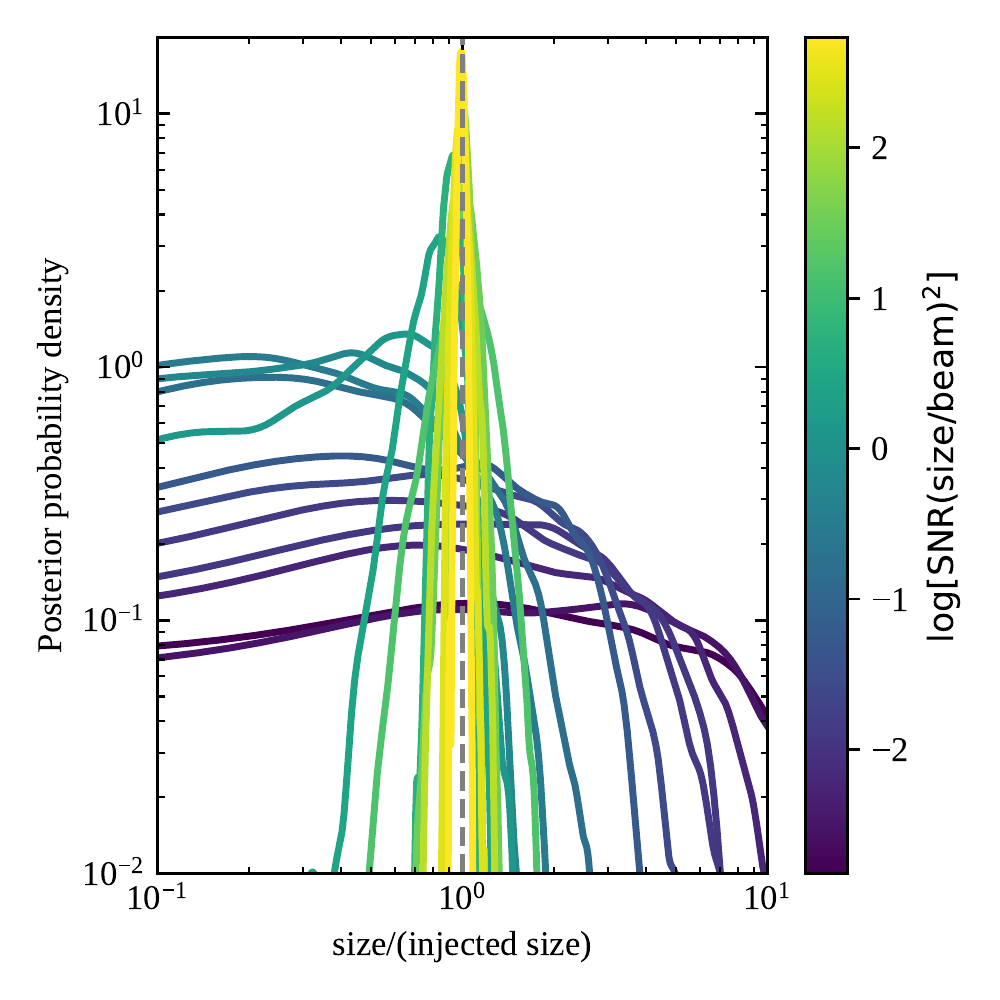}
 \caption{Marginalised source size posterior probability density for simulated sources immersed in real noise. Solid curves, colour-coded by the over-resolution parameter \citep{Marti-Vidal2012} $\omega=\mathrm{SNR}(\mathrm{size}/\mathrm{beam})^2$, show the marginalised posterior probability density of the source size (in units of the injected size) obtained by applying our MCMC-based fitting procedure to simulated circular Gaussian sources immersed in the noise of our Oct 03 VLBA 15 GHz observation. For $\omega\gtrsim 1$ the source starts to be clearly distinguishable from a point source.}
 \label{fig:simulfit}
\end{figure}
According to \citet{Marti-Vidal2012}, the possibility to over-resolve a source (i.e., being able to resolve it despite its size being smaller than the synthesised beam) depends critically on the parameter $\omega=\mathrm{SNR}(\mathrm{size}/\mathrm{beam})^2$ (see their Eq.~7), with $\omega\sim 1$ being the threshold below which the source cannot be resolved (the exact threshold depends on the array characteristics). Figure~\ref{fig:simulfit} reports the results of our simulations, showing how the marginalised posterior probability density of the source size depends on the $\omega$ over-resolution parameter. Our results are in excellent agreement with those of \citet{Marti-Vidal2012}, and they show that our chosen priors are well-behaved and that the analysis leads to unbiased results.

\subsection{Relativistic blastwave source size and Sedov length constraint}
\label{sec:sedov_length_constraint_method}
Our VLBI size measurements are obtained assuming a circular Gaussian source visibility model. In order to compare these to the expected size of a relativistic blastwave, we investigated the visibility amplitude dependence on the $UV$ radius for the brightness profile from \citet{Granot1999}, which is a limb-brightened disk whose physical radius is given by $R_\mathrm{\perp,max}=3.9\times 10^{16}(E_{52}/n)^{1/8}(t_\mathrm{obs}/(1+z)\mathrm{days})^{5/8}\,\mathrm{cm}$. We find the shape of the first peak in the visibility amplitudes to be essentially independent of the similarity variable \citep{Granot1999} $\phi$ as long as $0.1<\phi<10$, which comfortably accommodates our observations. 
\begin{figure}[t]
    \centering
    \includegraphics[width=0.6\textwidth]{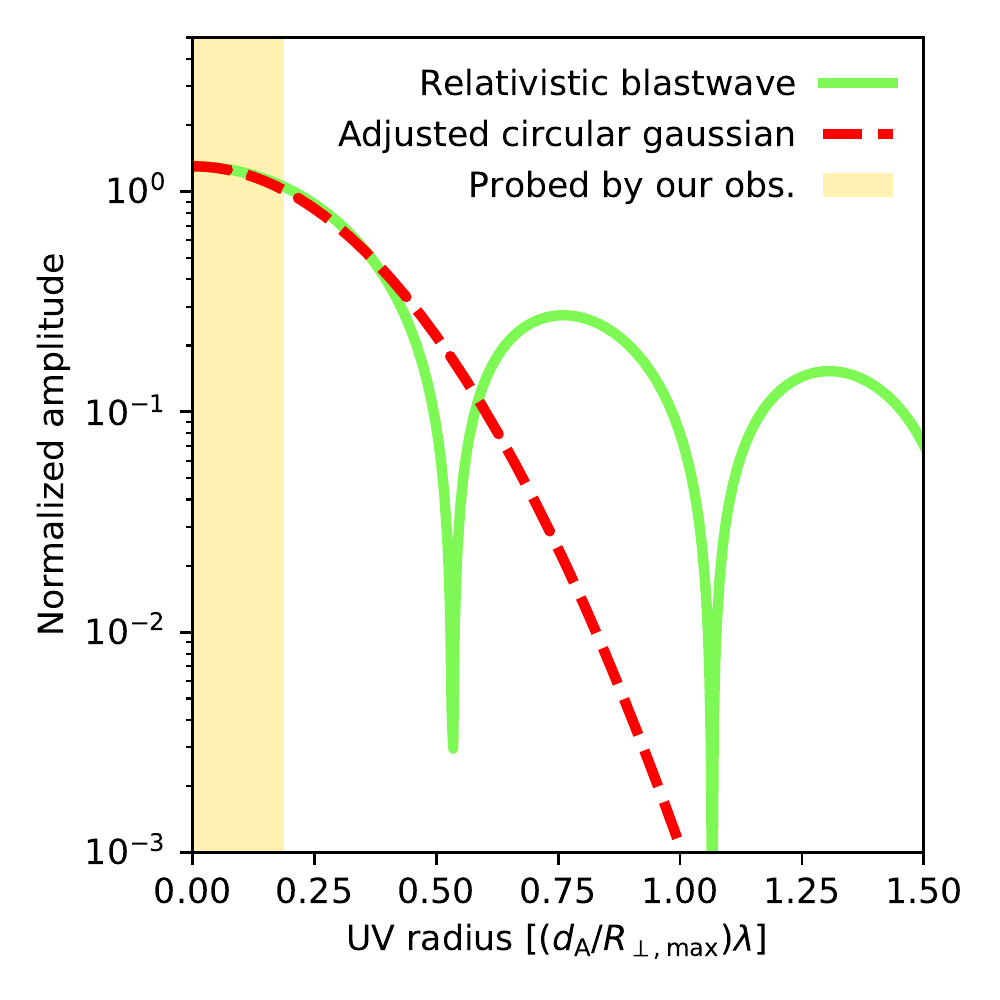}
    \caption{Comparison of the visibility amplitude dependence on the $UV$ radius for a circular Gaussian and a relativistic blastwave \citep{Granot1999}. The circular Gaussian size has been adjusted to match the first peak of the blastwave, see text. }
    \label{fig:blastwave_circ_gauss}
\end{figure}
In Fig.~\ref{fig:blastwave_circ_gauss} we plot the amplitude as a function of the $UV$ radius for $\phi=1$ (green solid line), along with that corresponding to a circular Gaussian (red dashed line) with a size (FWHM) $s=0.65\times 2 \times R_\mathrm{\perp,max}/d_\mathrm{A}$, where $d_\mathrm{A}$ is the angular diamter distance. This demonstrates that, as long as the longest baselines are shorter than $\sim 0.4 d_\mathrm{A}/R_\mathrm{\perp,max}$ wavelengths, such a circular Gaussian accurately reproduces the expected amplitudes. Our longest baselines extended to $\sim 160\,\mathrm{M}\lambda$, which is well below this limit at the times of our observations. This allowed us to make the identification $s=0.65 \times 2 \times R_\mathrm{\perp,max}/d_\mathrm{A}$, where $s$ is the FWHM of our circular Gaussian model. Inverting this, we obtained $\log(E/n)=52+8\log(s)-5\log(t_\mathrm{obs}/(1+z)\mathrm{days})-8\log(5.08\times 10^{16}\,\mathrm{cm}/d_\mathrm{A})$. This relation allowed us to turn the likelihood $\mathcal{L}(d_i\,|\,s,t_{\mathrm{obs},i})$ of the $i$-th observation ($d_i$ here represents the corresponding visibility dataset, $t_{\mathrm{obs},i}$ represents the time at which the observation was performed, and the likelihood is marginalised over all variables but the source size) into a $\log(E/n)$ likelihood. By Bayes' theorem, the posterior probability density is $P(\log(E/n)\,|\,d_i)\propto \mathcal{L}(d_i\,|\,\log(E/n),t_{\mathrm{obs},i})\pi(\log(E/n))$, where the last term is the prior on $\log(E/n)$. Our flat prior on the source size $s$ corresponds to a prior $\pi(\log(E/n))\propto (E/n)^{1/8}$. We note that in our afterglow modelling, given the chosen priors (see Table~\ref{tab:afterglow_params}), the effective prior on $x = \log(E/n) = \log(E)-\log(n)$ is instead
\begin{equation}
 \pi(x) \propto \left\lbrace\begin{array}{lr}
                                     x-x_\mathrm{min} & x_\mathrm{min}\leq x < (x_\mathrm{min}+x_\mathrm{max})/2\\
                                     x_\mathrm{max}-x & (x_\mathrm{min}+x_\mathrm{max})/2\leq x \leq x_\mathrm{max} \\
                                    \end{array}\right.
\end{equation}
where $x_\mathrm{max}=\log(E_\mathrm{max}/n_\mathrm{min})$, $x_\mathrm{min}=\log(E_\mathrm{min}/n_\mathrm{max})$. Here $E_\mathrm{min}$, $E_\mathrm{max}$, $n_\mathrm{min}$ and $n_\mathrm{max}$ represent the prior bounds on $E$ and $n$ as reported in Table~\ref{tab:afterglow_params}. For the comparison in Fig.~\ref{fig:E_n_constraint} therefore we divided the afterglow posterior by this prior and multiplied it by $(E/n)^{1/8}$, in order to keep the prior consistent in the comparison (the effect is anyway negligible, as the likelihood is strongly peaked).

Since the size measurements are independent, their likelihoods could be combined by multiplication, so that the posterior probability of $\log(E/n)$ from the entire dataset could be expressed as $P(\log(E/n)\,|\,\mathbf{d}) \propto \pi(\log(E/n))\Pi_{i=1}^{9}\mathcal{L}(d_i\,|\,\log(E/n),t_{\mathrm{obs},i})$. The resulting posterior probability densities are shown in Fig.~\ref{fig:E_n_constraint}. From the combined epochs we obtained $E/n < 10^{54.1} \,\mathrm{erg\,cm^{3}}$ at the 90\% credible level (we only report the upper limit, since the lower bound is entirely determined by our prior). The (much tighter, but more model-dependent) estimate of $E/n$ we obtained from the multi-wavelength modelling of the afterglow emission (red line in Fig.~\ref{fig:E_n_constraint}) in agreement with this upper limit.

\section{Prompt emission: \textit{Fermi}/GBM data reduction}\label{sec:GBM_reduction}
The prompt emission light curve of GRB~190829A shows the presence of two emission episodes (we refer hereafter to these as episode I and episode II, respectively), separated by $\sim$50 sec. 
We analysed the spectra of the two prompt emission episodes detected by \textit{Fermi}/GBM \citep{Meegan2009}.
Spectral data files and the corresponding latest response matrix files (rsp2) were obtained from the online HEASARC archive\footnote{\url{https://heasarc.gsfc.nasa.gov/W3Browse/fermi/fermigbrst.html}}. Spectra were extracted using the public software \textsc{gtburst}. We analysed the data of the three most illuminated NaI detectors with a viewing angle smaller than 60$^{\circ}$ (n6, n7, and n9) and the most illuminated BGO detector (b1). In particular, we selected the energy channels in the range 8–900 keV for NaI detectors, excluding the channels in the range 30--40 keV (because of the iodine K–edge at 33.17 keV) and 0.3--40 MeV for the BGO detector. We used inter-calibration factors among the detectors, scaled to the most illuminated NaI and free to vary within 30 per cent. To model the background, we manually selected time intervals before and after the burst and modelled them with a polynomial function whose order is automatically found by \textsc{gtburst}. The spectral analysis has been performed with the public software \textsc{xspec}~(v.~12.10.1f). We used the PG-Statistic, valid for Poisson data with a Gaussian background, in the fitting procedure. 
   
For episode I, we performed a time-integrated analysis from $-1.79$ to 10.5 seconds after GBM trigger and fitted the spectra with two models, namely a power law with an exponential cutoff, and the Band function (namely, two power laws smoothly connected at the peak through an exponential transition). We compared the models based on the Akaike information criterion \citep{Akaike1974} (AIC), finding that both fit the spectra equally well ($\Delta\mathrm{AIC}\leq 3$), but the $\beta$ parameter in the Band function fit has large uncertainties. We therefore considered the cut-off power law as the best-fitting model of episode I, with best-fitting parameters: $\alpha = -1.63 _{-0.08}^{+0.09}$, $E_\mathrm{c} = 380 _{-134}^{+318}$ keV and $F = 1.98_{-0.58}^{+0.07} \times 10^{-7} \,\mathrm{erg\,cm^{-2}\,s^{-1}}$, where $\alpha$ is the low-energy spectral index and $E_\mathrm{c}$ is the scale energy of the spectral cutoff, and F is the flux integrated in the energy range 10 keV -- 10 MeV. With this parameters, the peak of the $\nu F_\nu$ spectrum is at $E_\mathrm{p} = 139.7 _{-21.3}^{+57.1}$  keV and the isotropic equivalent energy is $E_\mathrm{iso} = 1.02 _{-0.12}^{+0.10} \times 10^{49}$ erg.
   
Also for episode II we performed a time-integrated analysis in the interval 47.04 -- 62.46 s with the same approach. The best-fitting model in this case is the Band function with $\alpha = -0.602 _{-0.358}^{+0.002}$, $E_\mathrm{p} = 11.30 _{-0.90}^{+0.39}$  keV, $\beta = -2.52 _{-0.02}^{+0.01}$ and $F = 7.56_{-0.11}^{+0.12} \times 10^{-7} \,\mathrm{erg\,cm^{-2}\,s^{-1}}$, where $E_\mathrm{p}$ is the peak photon energy of the $\nu F_\nu$ spectrum, $\alpha$ and $\beta$ are the low-energy and high-energy spectral indices, respectively. For the second episode, the isotropic equivalent energy is $E_\mathrm{iso} = 2.81 _{-0.15}^{+0.17} \times 10^{50}$ erg.
The results of the spectral analysis of the prompt emission are consistent with those previously published in the literature, e.g.\ \citet{Lesage2019,Hu2020,Fraija2020,Chand2020}.

\section{Afterglow: data reduction and modelling}

\subsection{Data collection from the literature}\label{sec:data_from_literature}
We constructed an extensive GRB~190829A afterglow dataset combining publicly available data, the results of our VLBI flux density measurements, and our own analysis of \textit{Swift}/UVOT data. We obtained the \textit{Swift}/XRT unabsorbed flux light curve shown in Fig.~\ref{fig:lightcurves} from the \textsc{Burst Analyzer} provided by the United Kingdom Swift Science Data Centre \citep{Evans2010}. The $r$-band optical data are from GTC observations, from which the host galaxy contribution has been subtracted, as described in \citet{Hu2020}. At times $0.1<t/\mathrm{days}<1$, a possible excess due to the underlying supernova could be present. The $u$-band data are the result of our own analysis of publicly available \textit{Swift}/UVOT data, described below. The radio data comprises ATCA and NOEMA measurements described in the main text, AMI-LA and MeerKAT data from \citet{Rhodes2020}, and our own flux densities as reported in Table~\ref{tab:fit} and shown with stars in Fig.~\ref{fig:lightcurves}. An estimated host galaxy contribution \citep{Rhodes2020} of $f_\mathrm{host,15.5\,GHz}=0.15\pm 0.1\,\mathrm{mJy}$ has been subtracted from AMI-LA data, and the uncertainty summed in quadrature. Data points that result in upper limits after this subtraction are not shown in Fig.~\ref{fig:lightcurves} for presentation purposes, but are included in the afterglow model fitting. Optical and ultraviolet data have been corrected for the Milky Way interstellar dust extinction \citep{Schlafly2016} assuming $E(B-V)=0.05$, and for the host galaxy extinction adopting a Small Magellanic Cloud extinction curve and $E(B-V)=1\pm 0.1$, following \citet{Chand2020}. The resulting systematic uncertainty in the flux density has been summed in quadrature to the flux density measurement errors.

\subsection{UVOT data reduction}\label{sec:uvot_data}
UVOT images taken with the \textit{u} filter were analysed with the public \textsc{HEASOFT (version 6.25)} software package. The most recent version of the calibration database was used. An ultraviolet candidate counterpart is detected 246 s after the BAT trigger at a position consistent with GRB~190829A. Photometry was performed within a circular source-extraction region of 3 arcsec in radius. The background was extracted from a circular region with a radius of about 20 arcsec, close to our target but
without contamination from other sources. We created the light curve (Fig.~\ref{fig:lightcurves}) of the UVOT data using the \texttt{uvotproduct} tool, combining subsequent exposures until a significance of at least 3 sigma is reached. To estimate the contamination from the host galaxy we stacked all the \textit{u}-band observations together with the tool \texttt{uvotimsum}. We performed photometry on this stacked image within three selected circular regions (of 3 arcsec in radius) only containing host galaxy emission, 
at a similar separation from the galactic nucleus ($\sim$9 arcsec) as GRB~190829A and along the galactic plane. The ultraviolet contribution of the host galaxy at the position of GRB~190829A was estimated as the mean flux density of these three regions, with a conservatively estimated uncertainty equal to the statistical and systematic errors plus the standard deviation of the three regions summed in quadrature. The resulting contaminant host galaxy flux density was then subtracted from the flux densities obtained through \texttt{uvotproduct}. Low-significance points at times $>1\,\mathrm{d}$ were considered as upper limits, given the tighter limits from GTC \citep{Hu2020}.

\subsection{Swift/XRT data reduction}\label{sec:swift_data}

\input{XRT_table}

In order to build the spectral energy distributions (SEDs) at the times of the HESS observations and check our model predictions, shown in Fig.~\ref{fig:SSC_SED}, we retrieved the XRT spectral files from the \textit{Swift}/XRT online archive \footnote{\url{https://www.swift.ac.uk/xrt\_spectra}}. We analysed the spectral files with the public software \textsc{xspec}~(v.~12.10.1f). We excluded the energy channels below 0.3\,keV and above 10\,keV. Each spectrum is modelled with an absorbed power law, using the Tuebingen-Boulder interstellar dust absorption model \citep{Wilms2000} available in \textsc{xspec}. In particular, we used the \texttt{tbabs} model for the Galactic absorption (using $N_{\rm H}= 0.056 \times 10^{22}$ cm$^{-2}$, \citealt{Kalberla2005}), and the \texttt{ztbabs} model for the host galaxy absorption, adopting the source redshift $z=0.0785$. The intrinsic $N_{\rm H}$ was fixed to the value obtained from the time-resolved analysis of late XRT data. Indeed, in the 0.3--10 keV energy range, the fitted values of $N_{\rm H}$ and of the spectral index are closely correlated: a larger value of $N_{\rm H}$ allows for a softer spectrum, and \textit{vice versa}, so that the net result of their combination is consistent with the observed spectrum. As a consequence, the intrinsic variations in the spectral index can be misinterpreted as variations of $N_{\rm H}$ when both these parameters are free to vary. Since no $N_{\rm H}$ variation is expected at the times we analysed, we performed a time-resolved spectral analysis of the XRT data up to $10^7$ s after the {\it Fermi}/BAT trigger by leaving both the host $N_{\rm H}$ and the photon index free. We found that, at late times (from $2.8 \times 10^4$ s onward), the $N_{\rm H}$ parameter does not evolve and remains constant around $N_{\rm H}= 1.16 \times 10^{22}$ cm$^{-2}$. We therefore fitted the XRT spectra shown in Fig.~\ref{fig:SSC_SED} assuming the above-mentioned value of the intrinsic $N_{\rm H}$ and leaving as free parameters the normalisation and the spectral index of the power law. The results of the spectral analysis of the XRT data are reported in Table~\ref{tab:XRTparams}. We note that the results of the spectral analysis are consistent with those previously published in the literature for similar integration times \citep{HESS2021}.

\subsection{Afterglow model}\label{sec:afterglow_model}

\subsubsection{Dynamics during the reverse shock crossing}
During the reverse shock crossing, we described the system as consisting of four regions separated by the forward shock, contact discontinuity and reverse shock, respectively. We assumed all hydrodynamic quantities in each region to be uniform, that is, we neglected the shock profiles. Region 1 is the unperturbed external medium, which we assumed to be cold and to have a uniform number density $n$. Region 2 is the shocked ambient medium; region 3 is the shocked jet material; region 4 is the unperturbed jet material. We assumed regions 2 and 3 to move with the same Lorentz factor $\Gamma$ during this phase, and we assumed the adiabatic index in both regions to be $\hat\gamma = 4/3$, that is, we assumed their pressure to be always radiation-dominated. Pressure balance across the contact discontinuity requires the internal energy densities in the two regions to be equal, namely $e'_\mathrm{int,2}=e'_\mathrm{int,3}$. Relativistic Rankine-Hugoniot jump conditions at the forward shock set $e'_\mathrm{int,2}=e'_\mathrm{int,3}=(\hat\gamma \Gamma + 1)(\Gamma-1)n m_\mathrm{p} c^2 /(\hat\gamma-1)$, where $m_\mathrm{p}$ is the proton mass, and $n'_\mathrm{e,2}=(\hat\gamma \Gamma + 1)n /(\hat\gamma-1)$. The number density in region 4 is given by $n'_\mathrm{e,4}=E_0/(4\pi R^2 \Delta'_4 \Gamma_0 m_\mathrm{p} c^2)$, where $E_0$ is the total, isotropic-equivalent jet kinetic energy, and $\Delta'_4(R) = \max\left[\Gamma_0 c T, R/\Gamma_0\right]$ where $T$ is the jet duration in the central engine frame, and we are assuming that the radial spreading of the jet becomes effective beyond the spreading radius $R_\mathrm{s}=\Gamma_0^2 c T$, after which the jet thickness in the central engine frame is \citep{Kobayashi2000numerical} $R/\Gamma_0^2$. Since the reverse shock is well-separated in time from the prompt emission, the shell was in the spreading phase at the time of deceleration, and the dynamics is therefore independent \citep{Kobayashi2000lightcurves} of $T$ (the so called `thin shell' regime). Shock jump conditions set $n'_\mathrm{e,3}=(\hat\gamma \Gamma_\mathrm{3,4} + 1)n /(\hat\gamma-1)$, where $\Gamma_\mathrm{3,4}=\Gamma_0\Gamma(1-\beta_0\beta)$, and $\beta_0=\sqrt{1-\Gamma_0^{-2}}$. The forward shock Lorentz factor, as measured in the central engine frame, is \citep{Blandford1976} $\Gamma_\mathrm{s,2}=((\Gamma+1)/(\hat\gamma(2-\hat\gamma)(\Gamma-1)+2))^{1/2}(\hat\gamma(\Gamma-1)+1)$. The same relation holds for the reverse shock Lorentz factor as measured in frame 3, changing $\Gamma$ with $\Gamma_{3,4}$. The reverse shock Lorentz factor $\Gamma_\mathrm{s,3}$ in the central engine frame was then obtained by the proper Lorentz transform. The amount of jet energy that crosses the reverse shock per unit radius advance is \citep{Nava2013} 
\begin{equation}
\frac{dE}{dR} = \frac{\beta_0-\beta_\mathrm{s,3}}{\beta_\mathrm{s,2}}\frac{E_0\Gamma_0}{\Delta'_4},    
\label{eq:en_crossing_rs}
\end{equation}
where $\beta_\mathrm{s,2}=\sqrt{1-\Gamma_\mathrm{s,2}^{-2}}$ and $\beta_\mathrm{s,3}=\sqrt{1-\Gamma_\mathrm{s,3}^{-2}}$, which can be integrated to give the jet energy that crossed the reverse shock at a given radius, $E(R)$. The comoving volume of regions 2 and 3 is $V'_i=4\pi R^2\Delta'_i$ with $i=2,3$. The thickness is set by electron (or baryon) number conservation, which yields $\Delta'_3 = (\hat\gamma-1)\Delta'_4 E(R)/(\hat\gamma\Gamma_{3,4}+1)\Gamma_0 E_0$ and $\Delta'_2 = (\hat\gamma-1)R /3(\hat\gamma\Gamma_{3,4}+1)$. The internal energy in region 2 is therefore $E_\mathrm{int,2}=e'_\mathrm{int,2}V'_2$ and similarly that in region 3 is $E_\mathrm{int,3}=e'_\mathrm{int,3}V'_3$. Finally, the mass swept by the forward shock is $m(R)=4\pi R^3 n m_\mathrm{p}/3$. All these relations allowed us to write the energy conservation equation
\begin{equation}
    E(R) + m(R)c^2 = \Gamma(R) E(R)/\Gamma_0 + \Gamma_\mathrm{eff}(R)\left[E_\mathrm{int,2}(R)+E_\mathrm{int,3}(R)\right]     
    \label{eq:en_cons_rs_fs}
\end{equation}
where $\Gamma_\mathrm{eff}=(\hat\gamma\Gamma^2-\hat\gamma+1)/\Gamma$ provides the proper transformation of the internal energies in the central engine rest frame \citep{Nava2013}. To compute the dynamical evolution, we started by assuming that the jet did not decelerate appreciably at a small initial radius $R_0=10^{10}\,\mathrm{cm}$, where we set $E(R_0)=0$, $\Gamma(R_0)=\Gamma_0$, $E_\mathrm{int,3}=0$ and $E_\mathrm{int,2}=(\Gamma_0-1)m(R_0)c^2$. We then iteratively advanced the radius by small logarithmic steps, solving numerically Eq.~\ref{eq:en_cons_rs_fs} for $\Gamma$ at each radius, and integrating Eq.~\ref{eq:en_crossing_rs} by the Euler method. 

To account for the effect of side expansion, we assumed regions separated by angular distances $\theta_\mathrm{s,0}=\Gamma_0^{-1}$ to be initially causally connected by pressure waves, and we assumed that the effective angle of causal connection $\theta_\mathrm{s}$ increases as
\begin{equation}
\frac{d\theta_\mathrm{s}}{dR} = \frac{\beta_\mathrm{sound}}{\beta\Gamma},    
\label{eq:dth_s/dR}
\end{equation}
where $\beta_\mathrm{sound}=\sqrt{\hat\gamma(\hat\gamma-1)(\Gamma-1)/(1+\hat\gamma(\Gamma-1))}$ is the proper sound speed behind the forward shock \citep{Kirk1999}. We integrated Eq.~\ref{eq:dth_s/dR} by the Euler method to obtain $\theta_\mathrm{s}(R)$. This computation proceeded from this phase into the subsequent phase after the reverse shock crossing is complete. We assumed the effective opening angle of the jet to be $\theta_\mathrm{j}(R)=\max\left[\theta_\mathrm{j}(R_0),\theta_\mathrm{s}(R)\right]$, that is, we assumed the jet to expand sideways at the local sound speed \citep{Huang1999,Lamb2018} as soon as the angular size of causally connected regions exceeded the initial angular size of the jet. The effect of side expansion on the dynamics is essentially that of diluting the jet energy over a larger solid angle, which we modelled simply by the substitution $E\to E(1-\cos\theta_\mathrm{j}(R_0))/(1-\cos\theta_\mathrm{j}(R))$ in Eq.~\ref{eq:en_cons_rs_fs}. We accounted for this also in computing the comoving number and energy densities that we used for the synchrotron emission modelling. 

\subsubsection{Dynamics after reverse shock crossing}
The condition $E=E_0$ marks the radius at which the reverse shock completely crosses the jet. In the thin shell regime (the relevant regime in our case), this happens approximately at the `decleration' radius
\begin{equation}
 R_\mathrm{dec} = \ell_S \Gamma_0^{-2/3}
 \label{eq:Rdec}
\end{equation}
where $\ell_S=(3E_0/4\pi\,n\,m_\mathrm{p}c^2)^{1/3}$ is the Sedov length. In the observer frame, this radius is crossed approximately at a time
\begin{equation}
 t_\mathrm{dec} = \frac{R_\mathrm{dec}}{2\Gamma_0^2 c} \sim 10^{-2} E_{0,53}^{1/3} n_{-1}^{-1/3} \Gamma_{0,2}^{-8/3}\,\mathrm{days}
 \label{eq:tdec}
\end{equation}
which corresponds to the peak time of the reverse shock emission.

From that radius on, we considered the evolution of the forward-shocked external medium material (region 2) as separated from that of the reverse-shocked jet material (region 3). In the thin shell regime (the relevant regime in our case \citep{Kobayashi2000lightcurves}) region 3 is expected to decelerate and expand adiabatically \citep{Kobayashi2000lightcurves}, transferring its energy \citep{Kobayashi2000numerical} to region 2, which continues its expansion in a self-similar manner \citep{Blandford1976}. We assumed \citep{Kobayashi2000lightcurves} region 3 to decelerate as $\Gamma_\mathrm{3}\propto R^{-g}$ and we adopted the usual value \citep{Kobayashi2000lightcurves} $g=2$, which is consistent with the results of relativistic hydrodynamical simulations \citep{Kobayashi2000numerical}. Adopting a polytropic equation of state $e'_3 \propto {n'}_\mathrm{e,3}^{\hat\gamma}$, the local sound speed is $c_\mathrm{s,3}\propto {n'}_\mathrm{e,3}^{(\hat\gamma-1)/2}$, which implies a comoving volume expansion $V'_3 \propto R^{(2g+6)/(\hat\gamma+1)}$. The internal energy therefore goes as $E'_\mathrm{int,3}\propto {V'}_3^{-1/3} \propto R^{-(2g+6)/(3\hat\gamma+3)}$ and the number density simply decreases as ${n'}_\mathrm{e,3}\propto {V'}_3^{-1}$. The initial conditions are given by the forward-reverse shock dynamics as computed in the previous section, and we kept the adiabatic index $\hat\gamma=4/3$ fixed throughout this phase. This completely describes the evolution of region 2 after the shock crossing.
For the evolution of region 2, we used a simplified energy conservation law \citep{Panaitescu2000}
\begin{equation}
    E_\mathrm{dyn} + mc^2 = \Gamma E_\mathrm{dyn}/\Gamma_0 + \Gamma^2 m c^2
    \label{eq:en_cons_fs}
\end{equation}
which can be solved analytically (therefore speeding up the computation) and gives rather accurate results down to the non-relativistic regime, despite the slightly incorrect transformation \citep{Nava2013} of the internal energy to the central engine frame. The `dynamical' isotropic-equivalent energy $E_\mathrm{dyn}$ here was defined as $E_\mathrm{dyn}=(E_0-E_\mathrm{3})(1-\cos\theta_\mathrm{j}(R_0))/(1-\cos\theta_\mathrm{j}(R))$, where $E_\mathrm{3}$ is the reverse-shocked material isotropic-equivalent total energy (except the rest-mass energy) in the central engine frame (the effects of side expansion were accounted for by the other factors in parentheses), that is, $E_\mathrm{3}=\Gamma_\mathrm{3,eff}E'_\mathrm{int,3} + (\Gamma_\mathrm{3}-1)E_\mathrm{0}/\Gamma_0$, where, again, $\Gamma_\mathrm{3,eff}=(\hat\gamma\Gamma_3^2-\hat\gamma+1)/\Gamma_3$ provides the correct relativistic transformation of the internal energy. This essentially means that we assumed all energy lost by the reverse-shocked material in this phase to be immediately transferred to the forward-shocked region, contributing to its expansion. For region 2, in this phase we accounted for the changing adiabatic index in the transition from the relativistic to the non-relativistic regime by adopting a simple fitting function \citep{Pe'er2012} $\hat\gamma=\hat\gamma(\Gamma)$ (reported in the cited article). This gives a more accurate estimate of the sound speed to be used in Eq.~\ref{eq:dth_s/dR} in this phase. 

\begin{figure}[t]
    \centering
    \includegraphics[width=0.6\textwidth]{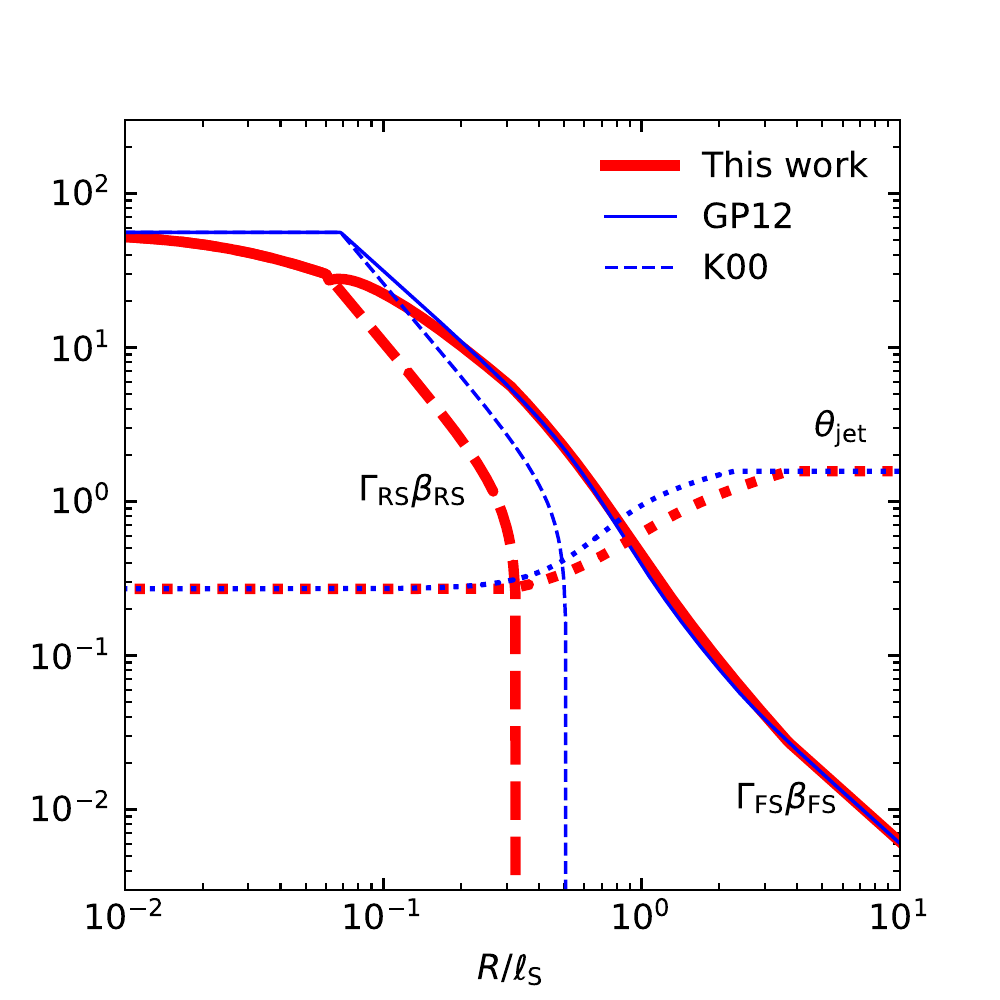} 
    \caption{Comparison between our shock dynamics and those from models in the literature. Red lines show the forward shock four-velocity (i.e.~$\Gamma\beta$, solid), reverse shock four-velocity (dashed) and jet half-opening angle (dotted) as a function of the shock radius (in units of the Sedov length) as computed with our model for our best fit parameters $E_0=2.5\times 10^{53}\,\mathrm{erg}$, $\Gamma_0=57$ and $n=0.21\,\mathrm{cm^{-3}}$. Blue lines show the same quantities computed following \citet[][their `trumpet' model]{Granot2012} for the forward shock Lorentz factor and jet opening angle, and \citet{Kobayashi2000lightcurves} for the reverse shock Lorentz factor.}
    \label{fig:dynamics_comparison}
\end{figure}

Fig.~\ref{fig:dynamics_comparison} shows a comparison of the forward shock dynamics computed with our model to that predicted by the `trumpet' model in \citet[][GP12 hereafter]{Granot2012}, and of the reverse shock dynamics compared to the simple analytical estimate from \citet[][K00 hereafter]{Kobayashi2000lightcurves}. The results are similar, though some differences are apparent: the side expansion predicted by our model is somewhat slower, and the initial deceleration is markedly different, since the GP12 model does not account for the reverse shock. 

\subsubsection{Computation of the light curves}
The dynamical computations described above give the Lorentz factor of the shock, $\Gamma_\mathrm{s}(R)$, and that of the shocked material, $\Gamma(R)$, as functions of the radius $R$ (the distance from the central engine) for each shock (forward and reverse), plus the effective comoving thickness $\Delta'(R)$ of the shocked region, the comoving internal energy density $e'_\mathrm{int}(R)$ and the electron number density $n'_\mathrm{e}(R)$ behind the shock. Given these quantities, we computed the comoving specific emissivity $j'_{\nu'}(R)$ behind the shock, by assuming (i) a fraction $\chi_\mathrm{e}$ of electrons to be accelerated into an isotropic, power-law energy distribution with index $p$, that is, $dn'_\mathrm{e}/d\gamma \propto \gamma^{-p}$ (where $\gamma$ is the electron Lorentz factor as measured in the comoving frame) extending from $\gamma_\mathrm{m}$ to $\gamma_\mathrm{max}$ and to hold a constant fraction $\epsilon_\mathrm{e}$ of the post-shock energy density at any radius, and (2) an effectively isotropic magnetic field $B$ to be generated by small-scale turbulence, again holding a constant fraction $\epsilon_\mathrm{B}$ of the post-shock energy density. These assumptions led to the definition of the injected electron power law minimum Lorentz factor
\begin{equation}
    \gamma_\mathrm{m} = \frac{1-\kappa^{1-p}}{1-\kappa^{2-p}}\frac{p-2}{p-1}\frac{\epsilon_\mathrm{e}}{\chi_\mathrm{e}}\frac{e'_\mathrm{int}}{n'_\mathrm{e}m_\mathrm{e}c^2}
\end{equation}
where $m_\mathrm{e}$ is the electron rest mass and $\kappa=\gamma_\mathrm{max}/\gamma_\mathrm{m}$ is taken as a free parameter. Whenever this value fell below $1$, we considered an effective injection Lorentz factor $\gamma_\mathrm{m,eff}= 1$ and an effective number of synchrotron-emitting electrons \citep{Sironi2013}  $\chi_\mathrm{e}n'_\mathrm{e,\,eff}=(1-\kappa^{1-p})(p-2)\epsilon_\mathrm{e}e'_\mathrm{int}/(1-\kappa^{2-p})(p-1)m_\mathrm{e}c^2$ (this is relevant early in the reverse shock in the thin shell regime,  and at late times in the forward shock, in the so-called `Deep Newtonian' phase, \citealt{Sironi2013}). The effective electron cooling Lorentz factor $\gamma_\mathrm{c}$ was computed as 
\begin{equation}
    \gamma_\mathrm{c} = \frac{6 \pi m_\mathrm{e} c}{\sigma_\mathrm{T} B^2 t' (1+Y)}
    \label{eq:gamma_c}
\end{equation}
where $\sigma_\mathrm{T}$ is the Thomson cross section, $t'$ is the comoving time elapsed since the explosion and $Y$ is the ratio of the comoving synchrotron radiation energy density. To account for the Klein-Nishina suppression of the cross section for photons with energy above $m_\mathrm{e}c^2$ in the electron comoving frame, we computed $Y$ including only the radiation energy density of photons below $\nu^\prime_\mathrm{KN}=m_\mathrm{e}c^2/h\gamma_\mathrm{c}$. This turned Eq.~\ref{eq:gamma_c} into an equation for the quantity $\gamma_\mathrm{c}(1+Y(\gamma_\mathrm{c}))$, which we solved numerically to obtain $\gamma_\mathrm{c}$ self-consistently. The electron energy distribution at a given radius, accounting for the effect of cooling, was thus assumed to have the form
\begin{equation}
    \frac{dn'_\mathrm{e}}{d\gamma} = \frac{\chi_\mathrm{e}n'_\mathrm{e}}{\gamma_\mathrm{p}\xi}\,\Xi(\gamma,\gamma_\mathrm{m},\gamma_\mathrm{c},p)
    \label{eq:dn/dgamma}
\end{equation}
where $\xi = \int_1^{\infty}\,\Xi\,d(\gamma/\gamma_\mathrm{p})$, and
\begin{equation}
    \Xi = \left\lbrace\begin{array}{lr}
    (\gamma/\gamma_\mathrm{p})^{-q} & \gamma_\mathrm{p}<\gamma<\gamma_0 \\
    (\gamma_0/\gamma_\mathrm{p})^{-q}(\gamma/\gamma_\mathrm{0})^{-p-1} & \gamma_0<\gamma<\gamma_\mathrm{max} \\
    0 & \mathrm{otherwise}
    \end{array}\right.
\end{equation}
where $\gamma_\mathrm{p}=\min(\gamma_\mathrm{m},\gamma_\mathrm{c})$, $\gamma_0=\max(\gamma_\mathrm{m},\gamma_\mathrm{c})$ and $q=2$ if $\gamma_\mathrm{c}<\gamma_\mathrm{m}$ or $q=p$ otherwise. The synchrotron emissivity of these electrons was assumed to be given by
\begin{equation}
    j'_\mathrm{\nu',syn} = \frac{\chi_\mathrm{e}n'_\mathrm{e} m_\mathrm{e} c^2 \sigma_\mathrm{T} B}{6 e \xi}S_\mathrm{\nu'}
\end{equation}
where $e$ is the electron charge, $\sigma_\mathrm{T}$ is the Thomson scattering cross section, and
\begin{equation}
\begin{array}{l}
    S_\mathrm{\nu'} = \left[C_1^s \left(\frac{\nu'}{\nu'_\mathrm{p}}\right)^{-s/3}+C_2^s \left(\frac{\nu'}{\nu'_\mathrm{p}}\right)^{s(q-1)/2}\right]^{-1/s}\times\\
    \times\left[1+\left(\frac{\nu'}{\nu'_\mathrm{0}}\right)^{1-q+p}\right]^{-1/2} \exp\left(-\sqrt{\frac{\nu'}{\nu'_\mathrm{max}}}\right)
\end{array}
\end{equation}
where $\nu'_\mathrm{p}=\gamma_\mathrm{p}^2 eB/2\pi m_\mathrm{e} c$, $\nu'_\mathrm{0}=\gamma_\mathrm{0}^2 eB/2\pi m_\mathrm{e} c$, $s=(q/3)^{-3/2}$, $C_1=5(q-1/3)/12$ and $C_2=(4/3)^q/2$. This is a fitting formula that approximates the exact spectral shape of synchrotron emission \citep{Rybicki1986} from the electron distribution in Eq.~\ref{eq:dn/dgamma}, and we include an exponential cut-off at the synchrotron burnoff \citep{DeJager1996} frequency $\nu'_\mathrm{max}=30\,\mathrm{MeV}$. Compared to the usual broken power-law approximation employed in the literature \citep{Sari1998,Granot1999,Panaitescu2000}, this gives a more accurate representation of the transitions between different spectral regimes. The synchrotron self-Compton emissivity was assumed to be
\begin{equation}
\begin{array}{l}
    j'_\mathrm{\nu',ssc} = \frac{m_\mathrm{e}c^2}{3e}\frac{\sigma_\mathrm{T}^2 \chi_\mathrm{e}^2 n^{'2}_\mathrm{e}}{4\pi \gamma_\mathrm{p}}B\frac{\zeta}{\xi^2 \lambda}\times \\
     \times \int_{\gamma_\mathrm{KN}}^{\gamma_\mathrm{max}} \Xi(\gamma/\gamma_\mathrm{p}) S_{\nu'}\left(\frac{3\nu'}{4\gamma^2}\right)d\gamma 
\end{array}
\end{equation}
where $\lambda=\int_0^{\infty}S_{\nu'}d\nu'$, $\zeta = \int_1^{\gamma_\mathrm{max}/\gamma_\mathrm{p}} (\gamma/\gamma_\mathrm{p})^2\, \Xi \,d(\gamma/\gamma_\mathrm{p})$, and
$\gamma_\mathrm{KN}=3h\nu'/4 m_\mathrm{e}c^2$ is the electron Lorentz factor below which Compton scattering at frequency $\nu'$ is suppressed by the Klein-Nishina effects. The total emissivity is $j'_{\nu'}=j'_\mathrm{\nu',syn} + j'_\mathrm{\nu',ssc}$.

For the reverse shock, during the adiabatic expansion phase that follows the shock crossing, we expect no more electrons to be injected into region 3, and no further acceleration to take place. Moreover, since the magnetic field energy density is thought to reach $\epsilon_\mathrm{B}e'_\mathrm{int,3}$ by means of amplification by turbulence behind the shock, it is reasonable \citep{Chang2008} to expect $\epsilon_\mathrm{B}$ not to remain constant after the shock crossing. For these reasons, we assumed the Lorentz factors $\gamma_\mathrm{m}$ and $\gamma_\mathrm{c}$ to evolve as $\gamma_\mathrm{x}(R)\propto \gamma_\mathrm{x}(R_\mathrm{\oplus})(V'_{3}(R)/V'_{3,\oplus})^{-1/3}$, where $x=\mathrm{m,c}$ and the $\oplus$ subscript denotes the quantities at the end of shock crossing, that is, we assumed the electron energy distribution evolution to be dominated by adiabatic cooling. We also neglected any emission from electrons above $\gamma_\mathrm{c}$, as no electrons above this energy are injected. Finally, we assumed the magnetic field to decay as $B_3(R) = B_{3,\oplus}(V'_{3}(R)/V'_{3,\oplus})^{-\eta_\mathrm{B}/2}$, where $\eta_\mathrm{B}$ is a constant that parametrizes our ignorance of the magnetic field decay in this phase. The frequency-dependent synchrotron self-absorption optical depth was computed as $\tau_\mathrm{\nu'}=\alpha_\mathrm{\nu'}\Delta'$. Here $\alpha_\mathrm{\nu'}$ is the appropriate absorption coefficient \citep{Rybicki1986}, which we decompose into $\alpha_\mathrm{\nu'}=a_0 a_\mathrm{\nu'}$, where 
\begin{equation}
    a_0 = 3^{((p+1)/2)}\left(\frac{1.8}{p^{0.7}}+\frac{p^2}{40}\right)\frac{(p-1)\pi^{3/2}e\,\chi_\mathrm{e} n'_\mathrm{e}}{\gamma_\mathrm{p}^5 B}
    \label{eq:ssa_abs_coeff}
\end{equation}
and
\begin{equation}
    a_\mathrm{\nu'} = \left\lbrace\begin{array}{lr}
        \left(\frac{\nu'}{\nu'_\mathrm{p}}\right)^{-5/3} & \nu'<\nu'_\mathrm{p} \\
        \left(\frac{\nu'}{\nu'_\mathrm{p}}\right)^{-(q+4)/2} & \nu'_\mathrm{p}<\nu'<\nu'_\mathrm{0} \\
        \left(\frac{\nu'_0}{\nu'_\mathrm{p}}\right)^{-(q+4)/2}\left(\frac{\nu'}{\nu'_\mathrm{0}}\right)^{-(p+5)/2} & \nu'_\mathrm{p}<\nu'<\nu'_\mathrm{0} 
    \end{array}\right.
\end{equation}
The dependence on $p$ in Eq.~\ref{eq:ssa_abs_coeff} is a fitting function \citep{Ghisellini2013} to the exact expression \citep{Rybicki1986}.
The comoving surface brightness at the shock was computed as $I'_{\nu'} = ((1-\exp(-\tau_\mathrm{\nu'}))/\tau_\mathrm{\nu'})j'_{\nu'}\Delta'(R)$. The surface brightness of the shock for an on-axis observer is then $I_\mathrm{\nu}(R) = (1+z)\delta^3 I'_\mathrm{(1+z)\nu/\delta}(R)$, where $\delta = \Gamma^{-1}(1-\beta\cos\theta)$, with $\beta=\sqrt{1-\Gamma^{-2}}$. Here $\theta$ is the polar coordinate of a reference frame centred at the central engine, whose $z$ axis coincides with the jet axis (and with the line of sight). In order to compute the light curves, we integrated such surface brightness over equal-arrival-time surfaces. To do so, we first computed the observer time 
\begin{equation}
t_\mathrm{obs}(R,\theta) = \frac{1}{c}\int_0^R (\beta_\mathrm{s}^{-1}-\cos\theta)dR,
\end{equation}
where $\beta_\mathrm{s}=\sqrt{1-\Gamma_\mathrm{s}^{-2}}$, on a grid over the jet surface. For computational efficiency, since most of the emission comes from regions that are closest to the line of sight (due to relativistic beaming), we used a logarithmically spaced grid in $\theta$, which provides finer spacing closer to the line of sight, and with the smallest grid spacing equal to $10^{-2}\Gamma_\mathrm{0}^{-1}$, where $\Gamma_\mathrm{0}$ is the initial jet Lorentz factor. This ensured the relativistic beaming cones were always resolved. We then numerically inverted the relation between $R$ and $t_\mathrm{obs}$ on each point of the grid, to obtain $R(t_\mathrm{obs},\theta)$, that is, the equal-arrival-time surfaces. The afterglow flux density was finally computed as
\begin{equation}
    \frac{dF}{d\nu} = \frac{2\pi}{d_\mathrm{L}^2} \int_0^{\theta_\mathrm{j}} R^2(t_\mathrm{obs},\theta) I_\mathrm{\nu}(R(t_\mathrm{obs},\theta))\sin\theta d\theta
\end{equation}
where $d_\mathrm{L}$ is the luminosity distance.

In Fig.~\ref{fig:size_evolution} we show the predicted size (full width at half maximum) of the radio image entailed by the model. This was computed using the model described in \citet{Ghirlanda2019}, which yields more accurate surface brightness distributions as it includes the integration over the shock profile.

\subsection{Photon-photon absorption optical depth}\label{sec:tau_gammagamma_afterglow} 
\begin{figure}[t]
    \centering
    \includegraphics[width=0.6\textwidth]{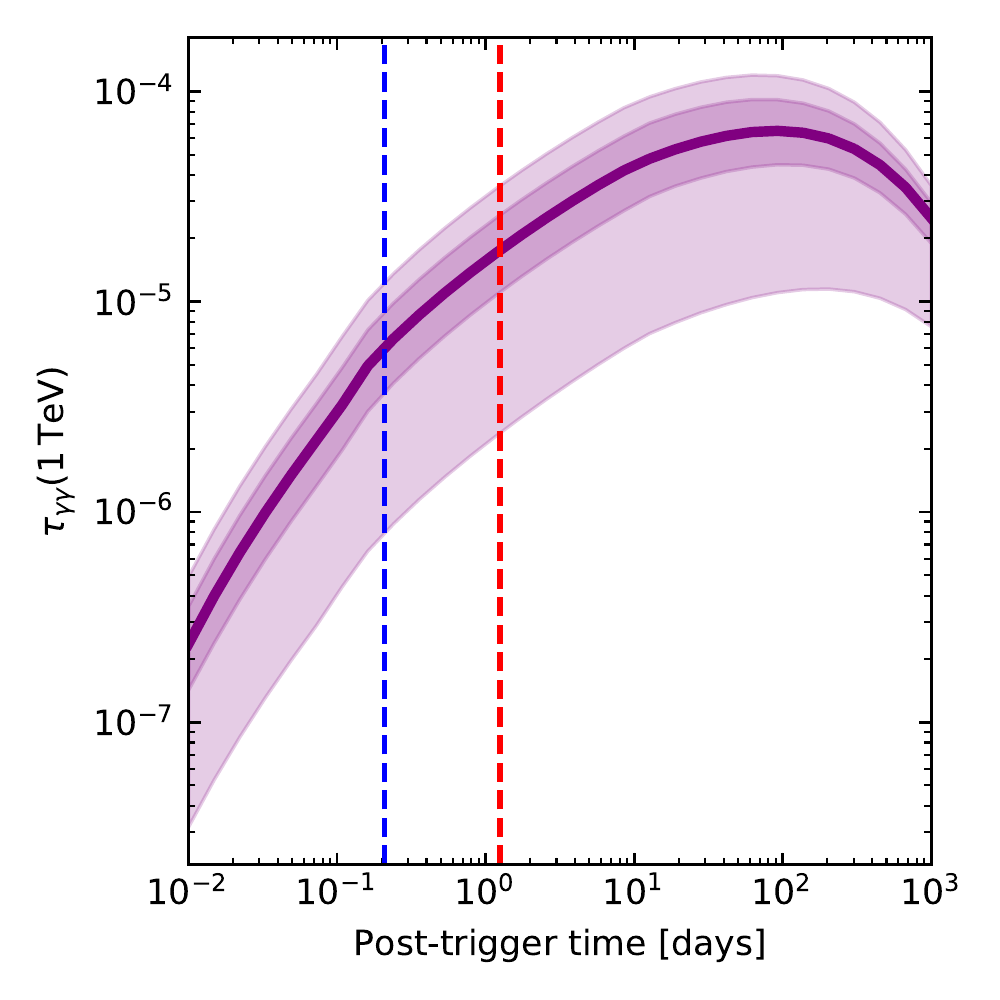}
    \caption{Photon-photon annihilation optical depth for 1 TeV photons. The purple line shows the optical depth to photon-photon annihilation for photons of 1 TeV observed energy produced in the forward shock downstream, as a function of the observed time after the gamma-ray trigger. The bands show the 50\% (darker band) and 90\% (lighter band) uncertainty propagated from the modelling uncertainties. The vertical dashed lines show the times of the HESS observations, namely 5 hours (blue) and 30 hours (red).}
    \label{fig:tau_gammagamma}
\end{figure}
High-energy photons produced in the shock downstream could have a non-negligible probability of pair-annihilate with lower energy photons, to form electron-positron pairs, before being able to escape the region. Therefore, we estimated the optical depth to this form of absorption for photons in the HESS energy range. The optical depth can be written approximately as \citep{Svensson1987}
\begin{equation}
    \tau_\mathrm{\gamma\gamma}(\nu') = \eta(\nu'_a) \sigma_\mathrm{T} \frac{dU_\mathrm{rad}(\nu'_a)}{dh\nu'}\Delta'
    \label{eq:tau_gammagamma}
\end{equation}
where $h$ is Planck's constant, $h \nu'_a=(m_\mathrm{e}c^2)^2/h\nu'$ is the typical frequency of target photons that can annihilate with those of frequency $\nu'$, $\eta(\nu'_a)$ is a dimensionless function \citep{Svensson1987} that depends on the slope of the photon spectrum at $\nu'_a$, $U_\mathrm{rad}$ is the comoving radiation energy density,  and $\Delta'$ is the comoving thickness of the shell. For photons in the HESS energy band, $\nu'_a$ is in the synchrotron range, so we can safely neglect the synchrotron self-Compton contribution to $U_\mathrm{rad}$. Also, we conservatively assume $\Delta'$ to be equal to the entire shell thickness as computed in the forward shock dynamics model described above, even though high-energy photons are mostly produced in a thinner shell closer to the shock, as fast enough electrons typically cool before being advected to the back of the shell. With these assumptions, we have that the optical depth for photons at observed frequency $\nu$ is
\begin{equation}
    \tau_\mathrm{\gamma\gamma}(\nu) = \eta(\nu'_a) \frac{m_\mathrm{e}c \sigma_\mathrm{T}^2}{9 eh}\frac{\chi_\mathrm{e} n'_\mathrm{e} B R^2 (\hat\gamma-1)}{\Gamma(\hat\gamma\Gamma+1)} \frac{\zeta}{\xi \lambda} S_\mathrm{\nu'}(\nu'_a)
\end{equation}
with $\nu'_a = \Gamma (m_\mathrm{e}c^2)^2/h^2\nu(1+z)$. Figure~\ref{fig:tau_gammagamma} shows the resulting optical depth for 1 TeV photons as a function of time, including the modelling uncertainties. The vertical dashed lines mark the times of the HESS observations. We conclude that photon-photon absorption is unimportant for our parameters.

\subsection{Afterglow model fitting}\label{sec:afterglow_model_fitting}

\begin{table*}[t]
\caption{Afterglow parameter estimation results. The columns report, from left to right, the parameter name and units, best fit value (maximum \textit{a posteriori}) with one-sigma errors (or 90\% upper/lower limits if the one-sigma credible interval rails against a lower/upper bound) for the narrow ($10^{-2}\leq \chi_\mathrm{e,FS}\leq 1$) and for the wide ($10^{-10}\leq \chi_\mathrm{e,FS}\leq 1$) prior, the bounds used in the MCMC fitting and the prior type adopted$^b$. The quantities below the horizontal line are derived from the fitting parameters.}
\centering
\begin{tabular}{lcccc}
Parameter$^a$ &  narrow prior & wide prior & bounds & prior type$^b$\\
\hline
$E_0/\mathrm{10^{53}\,erg}$ & $2.5^{+1.9}_{-1.3}$ & $8.6^{+26.0}_{-6.7}$ & $10^{48} - 10^{56}$ & l.u. \\
$n/\mathrm{cm^{-3}}$ & $0.21^{+0.37}_{-0.09}$ & $0.87^{+4.96}_{-0.63}$ & $10^{-6} - 10^{2}$ & l.u. \\
$\Gamma_0$ & $56.6^{+3.3}_{-5.3}$ & $55.0^{+3.4}_{-4.7}$ & $>10$ & l.~u.\\
$\theta_\mathrm{j}/\mathrm{deg}$ & $15.4^{+1.2}_{-0.94}$ & $15.8^{+1.3}_{-0.9}$ & $0.6 - 60$ & u.\\
$\epsilon_\mathrm{e,FS}$ & $0.030^{+0.029}_{-0.017}$ & $0.008^{+0.003}_{-0.006}$ & $10^{-6} - 0.6$ & l.~u.\\
$\epsilon_\mathrm{B,FS}/10^{-5}$ & $2.5^{+3.5}_{-1.3}$ & $<0.43$ & $10^{-6} - 0.3$ & l.~u.\\
$p_\mathrm{FS}$ & $2.010^{+0.002}_{-0.002}$ & $2.010^{+0.003}_{-0.002}$ & $2.001 - 2.9$ & u.\\
$\chi_{e,FS}/10^{-2}$ & $<6.5$ & $0.7^{+2.1}_{-0.5}$ & $10^{-2} (10^{-10}) - 10^{0}$ & l.u.\\
$\epsilon_\mathrm{e,RS}$ & $0.28^{+0.32}_{-0.16}$ & $0.1^{+0.5}_{-0.08}$ & $10^{-6} - 0.6$ & l.~u.\\
$\epsilon_\mathrm{B,RS}/10^{-3}$ & $1.2^{+1.8}_{-0.8}$ & $0.3^{+1.0}_{-0.2}$ & $10^{-6} - 0.3$ & l.~u.\\
$p_{RS}$ & $2.13^{+0.04}_{-0.08}$ & $2.12^{+0.05}_{-0.07}$ & $2.001 - 2.9$ & u.\\
$\log(\rho_{sys})$ & $-1.8^{+0.1}_{-0.1}$ & $-1.8^{+0.1}_{-0.1}$ & $10^{-10}-10^{0}$ & l.u. \\
\hline
$E_\mathrm{jet}/10^{51}\mathrm{erg}$ & ${9.4}_{-4.2}^{+8.9}$ & ${32}_{-25}^{+120}$ & -- & --\\
$\eta_\gamma/10^{-3}$ & ${1.1}_{-0.5}^{+1.2}$ & ${0.34}_{-0.26}^{+1.1}$ & -- & --\\
\end{tabular}
\flushleft
\footnotesize{
$^a$Credible ranges are computed as the smallest range that contains $68\%$ of the marginalised posterior probability, or 90\% for lower/upper limits.\\
$^b$l.~u. = 
log-uniform; u. = uniform}
    \label{tab:afterglow_params}
\end{table*}

\begin{figure*}
    \centering
    \includegraphics[width=\textwidth]{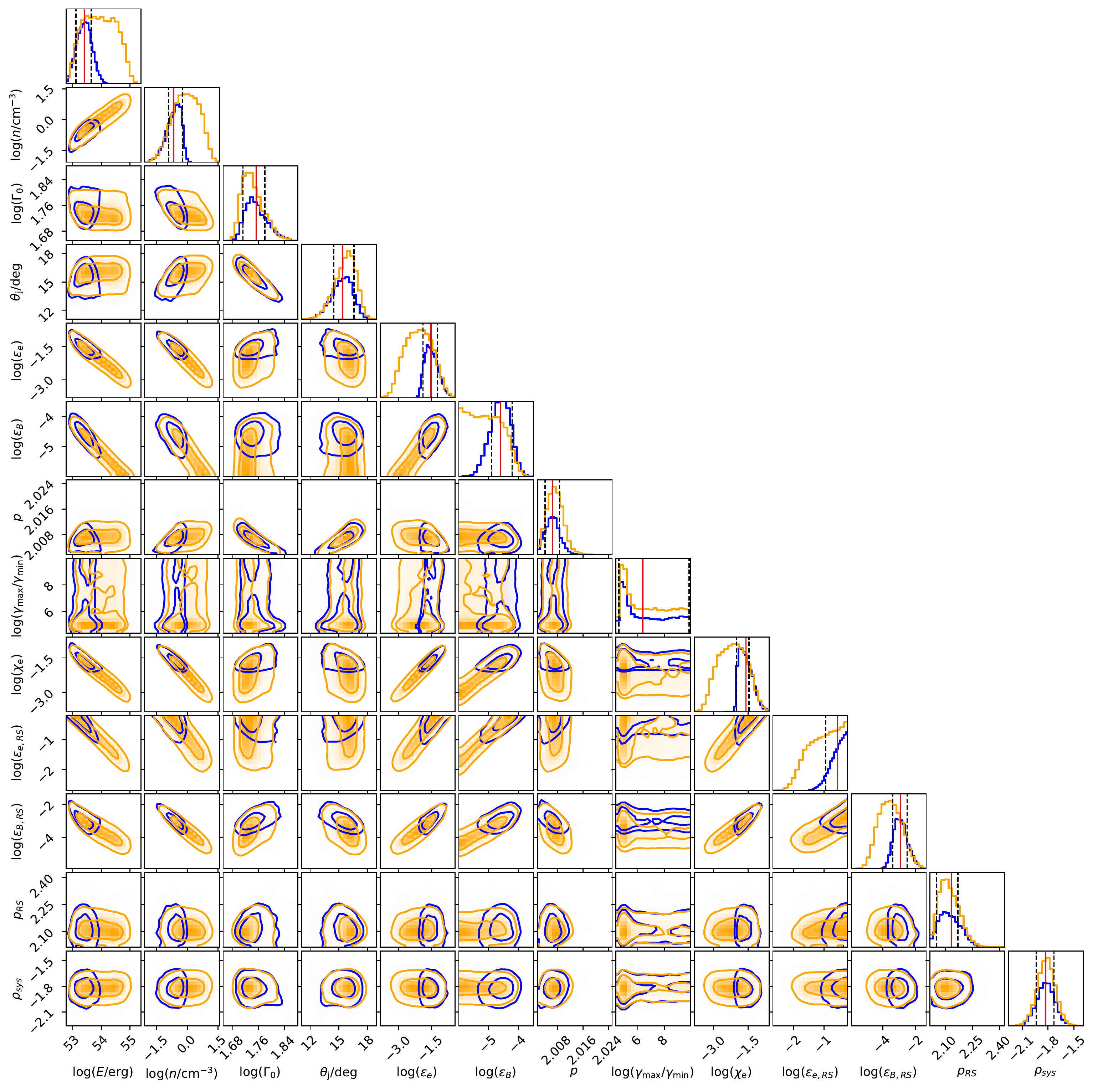}
    \caption{Corner plot of the Markov Chain Monte Carlo posterior sample density distributions. Yellow histograms and contours show the results for the wide $\chi_\mathrm{e}$ prior, while the blue ones are for the narrow prior. The black dashed lines in the plots on the diagonal bracket the one-sigma credible interval, while the red solid lines mark the position of the estimated best fit.}
    \label{fig:mcmc_total}
\end{figure*}

In order to estimate the parameters of our afterglow model that provide the best fit to the observations, and their uncertainties, we adopted an MCMC approach. We assumed a Gaussian log-likelihood model, to which each datapoint contributes an additive term
\begin{equation}
\begin{array}{l}
    \ln\mathcal{L}_\mathrm{i}(x) = -\frac{1}{2}\frac{\left(F_{\nu,\mathrm{m}}(x,\nu_\mathrm{i},t_\mathrm{obs,i})-F_\mathrm{\nu,i}\right)^2}{\sigma_\mathrm{i}^2+\rho_\mathrm{sys}^2 F_\mathrm{\nu,m}^2}+\\
    -\frac{1}{2}\ln\left[2\pi\left(\sigma_\mathrm{i}^2+\rho_\mathrm{sys}^2 F_\mathrm{\nu,m}^2\right)\right]
\end{array}
\end{equation}
where $F_\mathrm{\nu,i}$ is the $\mathrm{i}$-th flux density measurement, corresponding to frequency $\nu_\mathrm{i}$ and observer time $t_\mathrm{obs,i}$, or the flux integrated in the 0.3-10 keV band in the case of XRT datapoints (for these we also include the photon index, with a term of the same form but with no assumed systematic contribution to the uncertainty); $\sigma_\mathrm{i}$ is the associated one-sigma uncertainty (if asymmetric, the appropriate value is used depending on the sign of $F_{\nu,\mathrm{m}}-F_\mathrm{\nu,i}$). In the case of upper limits, we used a simple one-sided Gaussian penalty of the form 
\begin{equation}
\ln\mathcal{L}_\mathrm{i}=\left\lbrace\begin{array}{ll}
    0, &  F_{\nu,\mathrm{m}}-F_\mathrm{\nu,i}\leq 0\\
    -\frac{1}{2}\frac{\left(F_{\nu,\mathrm{m}}(x,\nu_\mathrm{i},t_\mathrm{obs,i})-F_\mathrm{\nu,i}\right)^2}{0.01^2F_\mathrm{\nu,i}^2}, &  F_{\nu,\mathrm{m}}-F_\mathrm{\nu,i}>0\\
\end{array}\right.
\end{equation}
The symbol $x$ here represents all emission model parameters, namely $x=(E_0$, $n$, $\Gamma_0$, $\theta_\mathrm{j}$, $\epsilon_\mathrm{e,FS}$, $\epsilon_\mathrm{B,FS}$, $p_\mathrm{FS}$, $\chi_\mathrm{e,FS}$, $\epsilon_\mathrm{e,RS}$, $\epsilon_\mathrm{B,RS}$, $p_\mathrm{RS}$, $\chi_\mathrm{e,RS}$, $\eta_\mathrm{B}$). The additional dimensionless parameter $\rho_\mathrm{sys}$ represents an unknown systematic contribution to the relative error on all measurements, which is introduced to account for inter-calibration uncertainties between different instruments and to avoid datapoints with very small formal errors to dominate the likelihood. We adopt a log-uniform prior $\pi(\rho_\mathrm{sys})\propto \rho_\mathrm{sys}^{-1}$ between $10^{-10}$ and $1$ and eventually marginalise over this parameter. Due to the very high dimensionality of the problem, and to the rather expensive computation of the likelihood (which requires the evaluation of the entire dynamics and emission model at a number of times and frequencies), we found keeping all parameters free to be intractable with our computational resources. By manual exploration of the parameter space we found that $\eta_\mathrm{B}<6$ always produced late bumps \citep{Resmi2016} in the radio band, which tended to overproduce the observed flux densities, while for $\eta_\mathrm{B}\geq 6$ these bumps were suppressed, so we fixed $\eta_\mathrm{B}=6$. We also found that the standard choice $\chi_\mathrm{e,RS}=1$ did not prevent a good fit of the early X-ray and optical data, so we kept also this parameter fixed. All other 12 parameters were left free to vary, subjected to the priors reported in Table~\ref{tab:afterglow_params}. We sampled the resulting posterior probability density using the \textsc{emcee} python package \citep{Foreman-Mackey2013}, using 24 walkers, which we run over $3\times10^{4}$ iterations, for a total of $7.2\times 10^5$ samples, on a cloud computing facility \citep{Landoni2018} provided by the Italian National Institute for Astrophysics (INAF). The resulting marginalised posterior probability density distributions, after discarding the initial 50\% of the chain as burn in, are shown in orange in the corner plot of Figure \ref{fig:mcmc_total}. 
After obtaining our solution, we verified that setting $\eta_\mathrm{B}<6$ significantly worsened the fit statistics when keeping all other parameters fixed to their best fit values.

\begin{figure}[t]
\centering
\includegraphics[trim=1cm 0cm 3cm 16.95cm,width=0.6\textwidth]{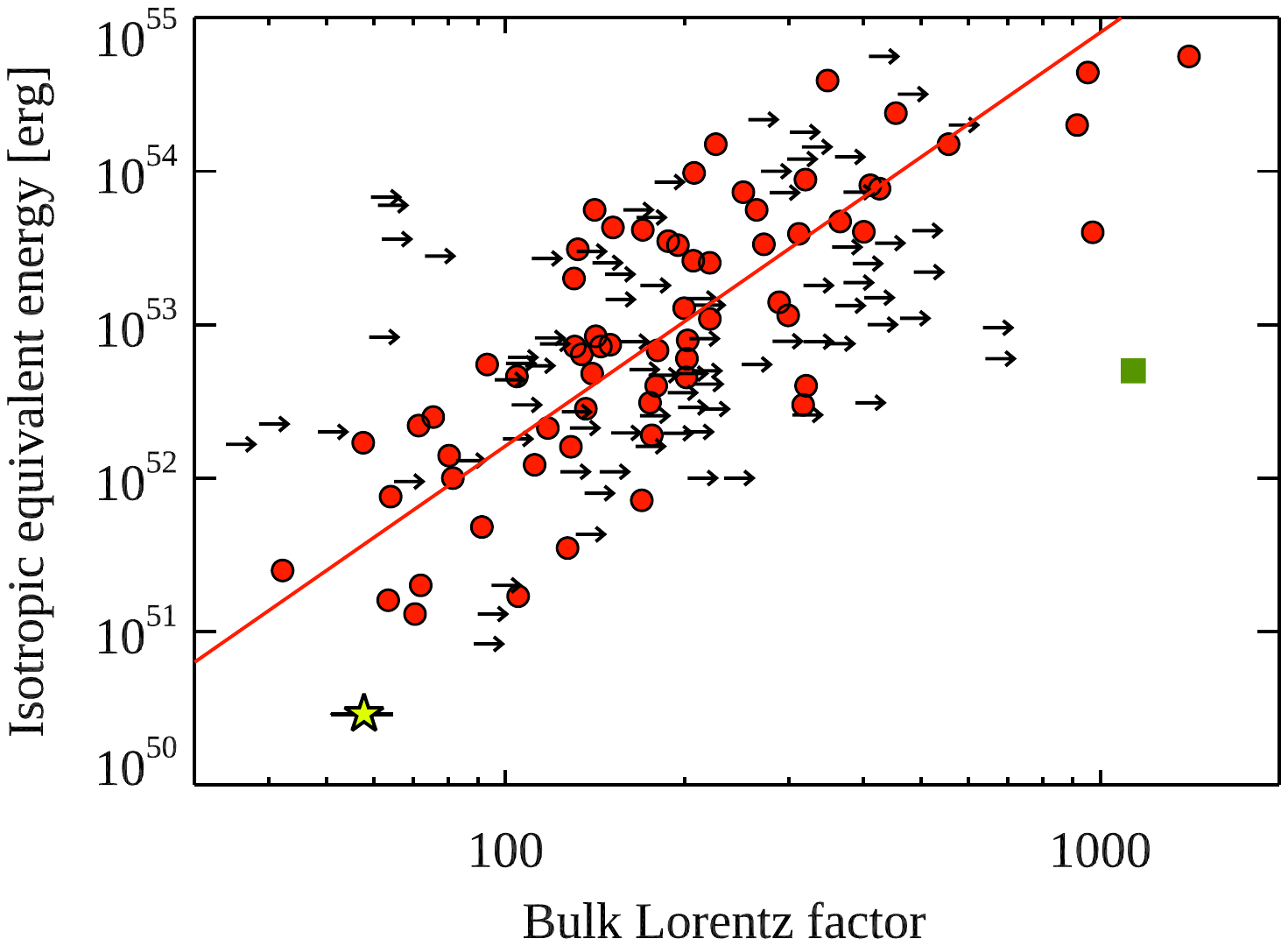}
\caption{Bulk Lorentz factor. Correlation between the gamma-ray isotropic equivalent energy and the bulk Lorentz factor for 67 long GRBs (red symbols) and one short GRB (green square). GRB~190829A is shown by the yellow star symbol. Lower limits on $\Gamma_{0}$ are also shown (black rightward arrows). Data are from \citet{Ghirlanda2018}. The solid line shows the power law that best fits the red points.}
\label{fig:eiso_gamma}
\end{figure}

Combining the bulk Lorentz factor estimate from this analysis and the isotropic equivalent energy from Appendix \ref{sec:GBM_reduction}, we can locate this burst on the $\Gamma - E_\mathrm{iso}$ plane. Figure~\ref{fig:eiso_gamma} compares the result with the sample from \citet{Ghirlanda2018}, which shows that this burst is consistent with the extrapolation of the previously observed correlation.

\bibliography{references}

\end{document}

%% file: Imaging-table.tex
\begin{table*}[t]
\caption{Summary of the VLBI imaging results of GRB~190829A. Columns give (1) epoch name as given in the main text, (2) modified Julian date (MJD), (3) observing frequency, 
(4) peak brightness, and the circular Gaussian model-fitting results: (5) total flux density, (6) 90\% credible upper limit on the angular size, {\it i.e.}, full width at half maximum, (7--8) relative offsets (including the correction of the core shift at 15~GHz of 0.035 mas in RA and $-$0.408 mas in Dec, see text) in Right Ascension and Declination with respect to the reference position (J2000, RA~=~$02^{\rm h}58^{\rm m}10^{\rm s}.52173$, Dec~=~$-08^\circ 57'28''.0956$). The reported uncertainties are statistical one-sigma errors; additional systematic uncertainties are discussed in the text.}
\label{tab:fit}
\centering \footnotesize
\begin{tabular}{cccccccc}
\hline\hline
Epoch & MJD       &  Freq  & $S_{\rm pk}$        & $S_{\rm tot}$     & $\theta_{\rm size,UL}$   & $\Delta$RA            &   $\Delta$Dec            \\
~ & (d)       &  (GHz) & (mJy~beam$^{-1}$)   & (mJy)             & (mas)                 &  (mas)                & (mas)                  \\
\hline
EVN (19 d) & 58743.947 &  4.99  & 0.546~$\pm$~0.017   & $0.645_{-0.016}^{+0.014}$ & $0.189$ & $-0.032_{-0.035}^{+0.028}$ & $-0.042_{-0.028}^{+0.021}$  \\ 
EVN (47 d) & 58771.880 &  4.99  & 0.340~$\pm$~0.011   & $0.352_{-0.011}^{+0.017}$ & $0.381$ & $0.080_{-0.047}^{+0.058}$ & $-0.004_{-0.054}^{+0.045}$  \\
EVN (75 d) & 58799.797 &  4.99  & 0.164~$\pm$~0.010   & $0.168_{-0.012}^{+0.016}$ & $1.331$ & $-0.245_{-0.090}^{+0.197}$ & $0.067_{-0.120}^{+0.158}$  \\
\hline
VLBA (9 d) & 58733.338 & 15.39  & 1.080~$\pm$~0.035   & $1.245_{-0.044}^{+0.064}$ & $0.120$ & $0.117_{-0.010}^{+0.015}$ & $-0.198_{-0.032}^{+0.032}$  \\ 
VLBA (35 d) & 58759.267 & 15.17  & 0.241~$\pm$~0.024   & $0.275_{-0.043}^{+0.033}$ & $0.285$ & $0.102_{-0.037}^{+0.053}$ & $-0.355_{-0.118}^{+0.118}$  \\ 
\hline
VLBA (13 d) & 58737.323 & 4.98  & 0.523~$\pm$~0.031   & $0.581_{-0.023}^{+0.019}$ & $0.477$ & $3.106_{-0.019}^{+0.018}$ & $3.378_{-0.058}^{+0.034}$  \\ 
VLBA (48 d) & 58772.238 & 4.98  & 0.229~$\pm$~0.017   & $0.248_{-0.014}^{+0.019}$ & $0.549$ & $1.500_{-0.044}^{+0.045}$ & $-0.314_{-0.120}^{+0.084}$  \\ 
VLBA (79 d) & 58804.140 & 4.98  & 0.128~$\pm$~0.010   & $0.105_{-0.012}^{+0.013}$ & $0.889$ & $1.032_{-0.109}^{+0.082}$ & $0.128_{-0.220}^{+0.231}$  \\ 
VLBA (116 d) & 58841.039 & 4.98  & 0.092~$\pm$~0.013   & $0.060_{-0.010}^{+0.030}$ & $3.198$ & $1.254_{-0.601}^{+0.198}$ & $-0.525_{-0.480}^{+0.399}$  \\ 

\hline

\end{tabular}

\end{table*}

%% file: XRT_table.tex
\begin{table*}[t]
\centering
\caption{Best fit parameters of the XRT data spectral analysis for the two SEDs epochs coincident with H.E.S.S. observations, as reported in Fig.~\ref{fig:SSC_SED}.}
\hspace{1.2cm}
\label{tab:XRTparams}
\begin{tabular}{cccc}
\hline\hline
Time interval & Flux @ 1 keV & Photon index & C-Stat/DOF \\
(h) & ($10^{-2}$ ph/ s cm$^2$ keV)  & & \\
\hline
$[4.89 - 7.85]$ & $1.60_{-0.07}^{+0.11}$ & $2.01_{-0.05}^{+0.04}$ & $398/420$ \\
$[29.02 - 29.16]$ & $0.38_{-0.09}^{+0.11}$ & $2.10_{-0.28}^{+0.23}$ & $43.5/280$ \\
\hline
\end{tabular}
\end{table*}
